\documentclass[11pt,draftcls,onecolumn] {IEEEtran}

\usepackage{setspace}		
\usepackage{cite}
\usepackage{array}
\usepackage{amssymb,amsmath}		

\usepackage{graphics}		
\usepackage{longtable}          
\usepackage{subfigure}

\usepackage{graphicx,array}
\newcommand{\vectornorm}[1]{{\left|\left|#1\right|\right|}_{\ell_2}} 
\newcommand{\vectornormOne}[1]{{\left|\left|#1\right|\right|}_{\ell_1}} 
\newcommand{\vectornormZero}[1]{{\left|\left|#1\right|\right|}_{\ell_0}} 

\newtheorem{theorem}{Theorem}[section]

\newtheorem{remark}[theorem]{Remark}

\newenvironment{proof}[1][Proof]{\begin{trivlist}
\item[\hskip \labelsep {\bfseries #1}]}{$\blacksquare$ \end{trivlist}}

\begin {document}

\title{Quantized Network Coding for Correlated Sources}
\author{Mahdy Nabaee,~\IEEEmembership{Student Member,~IEEE,}
        Fabrice Labeau,~\IEEEmembership{Member,~IEEE}
\thanks{M. Nabaee and F. Labeau are with the Department
of Electrical and Computer Engineering, McGill University, Montreal,
QC, E-mail: m.nabaee@ieee.org, fabrice.labeau@mcgill.ca}
\thanks{This work was supported by Hydro-Québec, the Natural Sciences and Engineering Research Council of Canada and McGill University in the framework of the NSERC/Hydro-Québec/McGill Industrial Research Chair in Interactive Information Infrastructure for the Power Grid.}
}
\maketitle

\begin{abstract}
\label{sec:abstract}
\boldmath
Non-adaptive joint source network coding of correlated sources is discussed in this paper.
By studying the information flow in the network, we propose quantized network coding as an alternative for packet forwarding.
This technique has both network coding and distributed source coding advantages, simultaneously.
Quantized network coding is a combination of random linear network coding in the (infinite) field of real numbers and quantization to cope with the limited capacity of links.
With the aid of the results in the literature of compressed sensing, we discuss theoretical and practical feasibility of quantized network coding in lossless networks.
We show that, due to the nature of the field it operates on, quantized network coding can provide good quality decoding at a sink node with the reception of a reduced number of packets.
Specifically, we discuss the required conditions on local network coding coefficients, by using restricted isometry property and suggest a design, which yields in appropriate linear measurements.
Finally, our simulation results show the achieved gain in terms of delivery delay, compared to conventional routing based packet forwarding.
\end{abstract}


\section{Introduction}
\label{sec:intro}

Flexible, low cost, and long lasting implementation of wireless sensor networks has made them an unavoidable alternative for conventional wired sensing structures in a wide variety of applications, including medicine, transportation, and military \cite{akyildiz2002survey}.
As a relatively new technology, the trends and challenges are more felt in the networking aspects of communication than in the classic physical layer era \cite{chong2003sensor}.
One of the introduced challenges is the gathering of sensed data at a central node of the network, where delivery delay, precision, and robustness to network changes are emerging issues.

As the conventional way of transmission in the networks, packet forwarding via routing is widely used in different implementations of sensor networks.
While it achieves capacity rates in the case of lossless networks \cite{netInfFlow}, packet forwarding requires an appropriate routing \cite{al2004routing} protocol to be run.
In the case of correlated sources, distributed source coding \cite{slepian1973noiseless,xiong2004distributed} on top of packet forwarding is proved to be optimal, in terms of achieved conditional information rates \cite{SWCtheorem}.
However, packet forwarding can lead to difficulties because of its need for queuing, and its slow adaptation to network changes, caused by deploying new node(s) or link failure(s).

These issues and a lot more have motivated the invention of network coding \cite{netInfFlow}, as an alternative for packet forwarding in sensor networks \cite{1228459,fragouli2009network}.
Specifically, network coding sends a function of incoming packets to the intermediate nodes, as opposed to sending their original content.
Furthermore, the usage of random linear functions, also known as random linear network coding, is proved to be sufficient in lossless networks \cite{koetter2003algebraic,NC_RLNCtoMulticast}.
Moreover, theoretical analysis shows that when network coding is used for transmission, no queuing is required to achieve the optimal information rates \cite{netInfFlow}.
Network coding in lossy networks can result in improved achieved rate regions, compared to packet forwarding \cite{lim2011noisy,erasNetCap}. 

Similar to packet forwarding, network coding can be separately applied on top of distributed source coding for correlated sources \cite{ho2004network,NCCorr_SepSCNC}.
On the other hand, one has to perform joint source network decoding in order to achieve optimal performance limits, which may not be feasible because of its computational complexity \cite{NCCorr_SepSCNC}. 
Sub-optimal solutions have been proposed to tackle this practicality issue \cite{wu2009practical,maierbacher2009practical,cruz2011joint}, by using low density codes and sum product algorithm \cite{kschischang2001factor} for decoding.
Similar to the distributed source coding, which requires knowledge of appropriate marginal rates at each encoder node, these approaches need some knowledge of correlation model of sources, at the encoders' side.
This knowledge of appropriate rates may be a luxury in some cases, especially when it is changing over time and needs to be updated frequently.
Hence, it is essential to study the possibility of developing a \textit{non-adaptive} joint source network coding for such cases.
In this paper, we aim to develop a non-adaptive random linear network coding for efficient joint distributed source network coding of correlated sources in sensor networks.

Recently, the idea of using compressed sensing \cite{CS,CSbook} and sparse recovery concepts in sensor networks has drawn attention \cite{rabbat,netcompass,CdataGathering,feizi2010compressive}.
For instance, in \cite{xu2011compressive,wang2012sparse}, theoretical discussion on sparse recovery of graph constrained measurements with an interest in network monitoring application is presented.
Joint source, channel and network coding was also proposed in \cite{feizi2011power}, where random linear mixing was proposed for compression of temporally and spatially correlated sources.
In \cite{bassi2012compressive}, practical possibility of finite field network coding of highly correlated sources was investigated, with the aid of low density codes and belief propagation base decoding.
Unfortunately, a solid theoretical investigation on the feasibility of adopting sparse recovery in random linear network coding has not been done previously.

In our earlier work \cite{naba1}, we proposed non-adaptive joint source network coding of exactly sparse sources, with the aid of the results in compressed sensing literature.
In this paper, we extend our work to the general case of correlated sources and discuss theoretical and practical aspects of having robust distributed compression.

A detailed description of data gathering scenario with our notations is presented in section~\ref{sec:probDescNotation}. In section~\ref{sec:QNC}, we introduce and formulate our proposed quantized network coding, which is followed by its theoretical feasibility discussion using restricted isometry property, in section~\ref{sec:RIP}.
In section~\ref{sec:L1Dec}, we present the decoding algorithm used to recover quantized network coded packets, and derive a performance bound on its recovery error.
Our simulation setup and results are presented in section~\ref{sec:SimRes}.
Finally in section~\ref{sec:conclusions}, we conclude the paper by discussion on our proposed method and the ongoing works on this topic.

\section{Problem Description and Notation}
\label{sec:probDescNotation}

In this paper, we consider a lossless model of networks, for which the links have limited capacities. Although it may not be a perfect model in practical cases where the links have mutual interference, it still reflects the effect of such imperfectnesses when calculating single input single output capacity of the links.
A future work may study the case of noisy networks of links, by understanding the effect of interference between the links.

\begin{figure}[t]
\centering
\resizebox{.38\textwidth}{!}{
\includegraphics[]{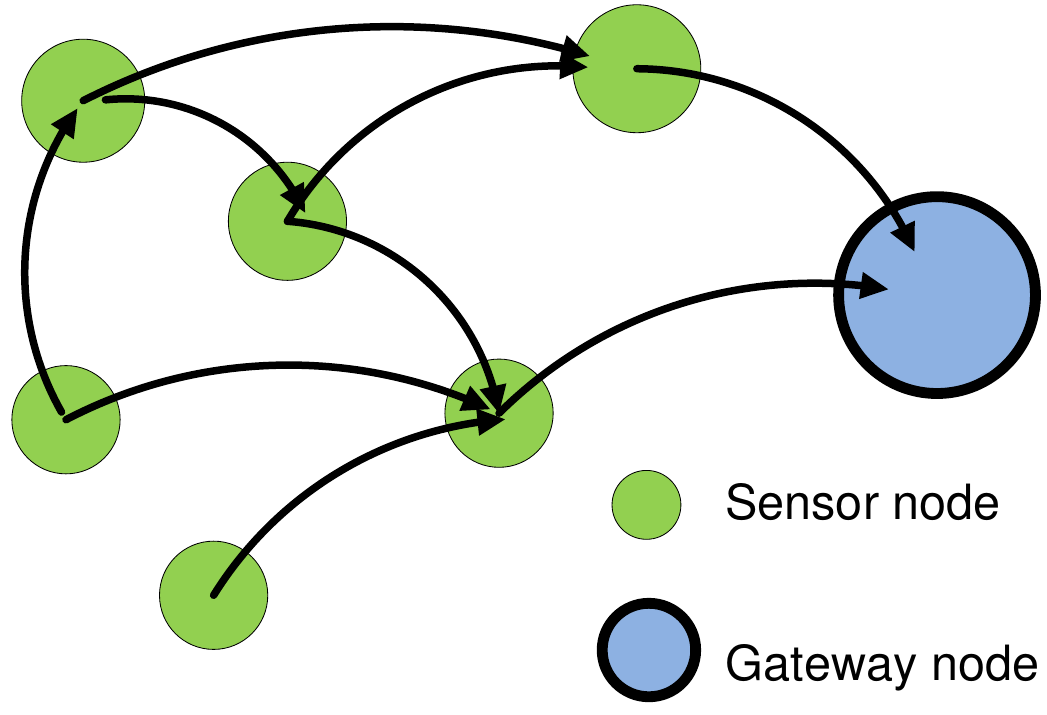}}
\caption[]{Directed graph, representing a data gathering sensor network.\label{fig:netDep}}
\end{figure}

As shown in Fig.~\ref{fig:netDep}, we represent the network by a directed graph, $\mathcal{G}=(\mathcal{V},\mathcal{E})$, where $\mathcal{V}$ and $\mathcal{E}$ are the sets of nodes (vertices) and directed edges (links).
Each node, $v$, is from the finite sorted set $\mathcal{V}=\{1,\cdots,n\}$ and each edge, $e$, is from the finite sorted set $\mathcal{E}=\{1,\cdots,|\mathcal{E}|\}$.
Further, each edge (link) can maintain a lossless transmission from $tail(e)$ to $head(e)$, at a maximum finite rate of $C_e$ bits per use.
We define the sets of incoming and outgoing edges of node $v$, denoted by $In(v)$ and $Out(v)$, respectively, as follows:
\begin{eqnarray}
In(v) &=& \{e:e \in \mathcal{E},~head(e)=v\}, \\
Out(v) &=& \{e:e \in \mathcal{E},~tail(e)=v\}.
\end{eqnarray}
The input and output contents of edge $e$ at time instant $t$ are represented by $Y_e(t)$ and $Y'_{e}(t)$, where $t$ represents the discrete (integer) time index, during which a block of length $L$ is transmitted.
Since the edges are lossless, $Y_e(t)$ and $Y'_{e}(t)$ are the same and from a finite alphabet of size $\lfloor 2^{L C_e} \rfloor$, where $\lfloor \centerdot \rfloor$ denotes truncation to the lower integer.
In the rest of the paper, the realizations of all capital letter random variables are denoted by lower case letters.

The nodes of the network are equipped with sensors and specifically each node $v$ has an information source, $X_v$, where $X_v \in \mathbb{R}$.
The sensed data are supposed to be correlated, as this is a valid assumption in a lot of different applications.
We model the correlation between these sensed data, by the near-sparseness property, since it can be considered as a generalization of compressibility and sparseness.
Specifically, by defining the sorted vector of $X_v$'s:
\begin{equation}
\underline{X}=[X_v: v \in \mathcal{V}], 
\end{equation}
we assume that $\underline{X}$ is near-sparse in some orthonormal transform domain $\phi_{n \times n}$.\footnote{In this paper, all the vectors are column-wise.}
Explicitly, for $\underline{S}=\phi^T \cdot \underline{X}$, and a small positive $\epsilon_k$, we have:
\begin{equation}\label{Eq:defEpsK}
\vectornormOne{\underline{S}-\underline{S}_k} \leq \epsilon_k,
\end{equation}
where $\underline{S}_k$ is such that:
\begin{equation}
\vectornormZero{\underline{S}_k}=k,
\end{equation}
and is called $k$-sparse.
An example of the sparsifying transform matrix, $\phi$, is the Karhunen Loeve transform of the messages.

Having these correlated information sources and the information network characterized, we study the transmission of $X_v$'s to a single gateway node.
The gateway or decoder node, denoted by $v_0$, $v_0 \in \mathcal{V}$, has high computational resources and is usually in charge of forwarding the information to a next level network; \textit{e.g.} a wired backbone network.
The described (single session) incast of sources to the unique decoder node is referred to as \textit{data gathering}.
The purpose of this paper is to discuss the theoretical and practical feasibility of \textit{non-adaptive} joint source network coding in the described data gathering scenario. More specifically, we take a compressed sensing approach in order to handle the transmission of sensed data.

\section{Quantized Network Coding}
\label{sec:QNC}
Random linear network coding for multicast of independent sources has been proposed and studied in \cite{NC_RLNCtoMulticast}, where the algebraic operations are in finite field.
Since our work is motivated by the concepts of compressed sensing, in which the results are valid in the infinite field of real number, we have to use a real field alternative for conventional finite field network coding.
On the other hand, finite capacity of the edges has to be appropriately coped with the infinite number of symbols in the adopted real field network coding.
As a result, we propose \textit{Quantized Network Coding} (QNC), which uses quantization to bridge between the limited capacity of the links and infinite alphabet of real field network coded packets.

\begin{figure}[t]
\centering
\resizebox{.45\textwidth}{!}{
\includegraphics[]{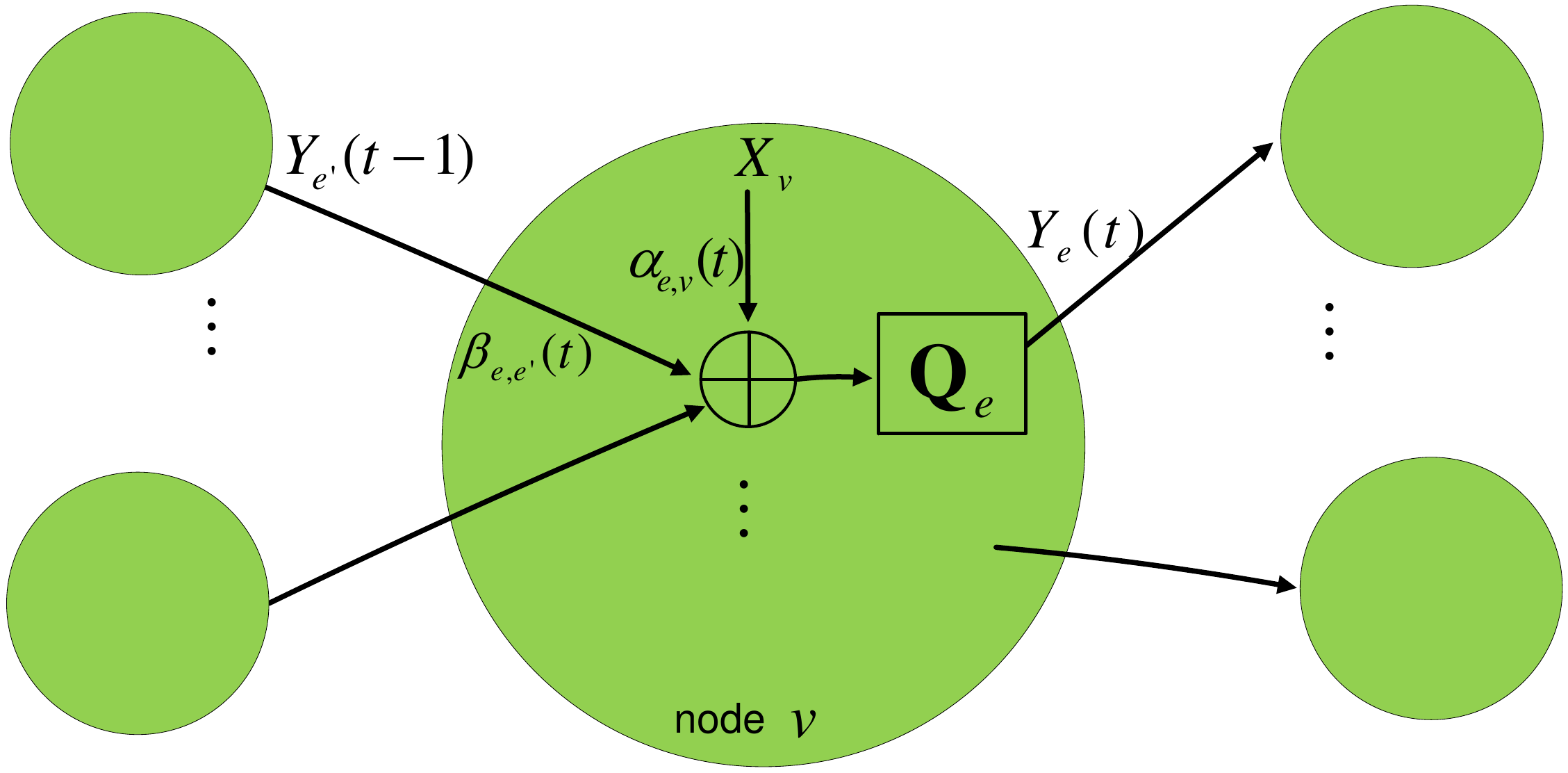}}
\caption[]{A simple diagram of quantized network coding.\label{fig:QNCdef}}
\end{figure}

In \cite{naba1}, for $\forall v \in \mathcal{V}, \forall e \in Out(v)$, we defined QNC at node $v$, according to:
\begin{equation}\label{Eq:QNCdef1}
Y_e(t)=\textbf{Q}_{e} \Big (\sum_{e' \in In(v)} \beta_{e,e'}(t) \cdot Y_{e'}(t-1)+\alpha_{e,v}(t) \cdot X_v \Big),
\end{equation}
where $Y_e(1)=0,~\forall e \in \mathcal{E}$, ensures initial rest condition in the network.
The messages, $X_v$'s are assumed to be constant until the transmission is complete.\footnote{This is why $X_v$'s do not depend on $t$.}
The local network coding coefficients, $\beta_{e,e'}(t)$'s and $\alpha_{e,v}(t)$'s are real valued and are usually picked semi-randomly. 
The quantizer operator, $\textbf{Q}_{e}(\centerdot)$, corresponding to outgoing edge $e$, is designed based on the values of $C_e$ and $L$, and the distribution of its input (\textit{i.e.} random linear combinations).
A simple diagram of QNC at node $v$ is shown in Fig.~\ref{fig:QNCdef}.

Denoting the quantization noise of $\textbf{Q}_{e}(\centerdot)$ at time $t$, by $N_e(t)$, we can reformulate (\ref{Eq:QNCdef1}) as follows:
\begin{equation}\label{Eq:QNCdef2}
Y_e(t)= \sum_{e' \in In(v)} \beta_{e,e'}(t) \cdot Y_{e'}(t-1)+\alpha_{e,v}(t) \cdot X_v +N_e(t).
\end{equation}
We denote the adjacency matrix, $[F(t)]_{|\mathcal{E}| \times |\mathcal{E}|}$, and $[A(t)]_{|\mathcal{E}| \times n}$ matrix, such that:
\begin{equation}\label{Eq:defF}
\{F(t)\}_{e,e'}=\left\{
\begin{array}{l l}
  \beta_{e,e'}(t)  & ,~tail(e)=head(e') \\
  0  &  ,~\mbox{otherwise} \\ \end{array} \right.,
\end{equation}
\begin{equation}\label{Eq:defineAt}
\{A(t)\}_{e,v}=\left\{
\begin{array}{l l}
  \alpha_{e,v}(t)  & ,~tail(e)=v \\
  0  &  ,~\mbox{otherwise} \\ \end{array} \right..
\end{equation}
We also define the vectors of edge contents, $\underline{Y}(t)$, and quantization noises, $\underline{N}(t)$, according to:
\begin{eqnarray}
\underline{Y}(t) & = & [Y_e(t):e \in \mathcal{E}], \\
\underline{N}(t) & = & [N_e(t):e \in \mathcal{E}].
\end{eqnarray}
As a result, Eq.~\ref{Eq:QNCdef2} can be re-written in the following form:
\begin{equation}\label{Eq:matrixForm}
\underline{Y}(t)=F(t) \cdot \underline{Y}(t-1)+A(t) \cdot \underline{X}+\underline{N}(t).
\end{equation}
Depending on the network deployment, matrix $[B]_{|In(v_0)| \times |\mathcal{E}|}$ defines the relation between the content of edges, $\underline{Y}(t)$, and the received packets at the decoder node.
Explicitly, we define the vector of \textit{marginal measurements} (received packets) at time $t$ at the decoder:
\begin{equation}
\underline{Z}(t)=[Y_e(t): e \in In(v_0)]=B \cdot \underline{Y}(t),
\end{equation}
where:
\begin{equation}
\{B\}_{i,e}=\left\{
\begin{array}{l l}
  1  & ,~i~\mbox{corresponds to}~e,~e \in \textit{In}(v_0) \\
  0  &  ,~\mbox{otherwise} \\ \end{array} \right..
\end{equation}

By considering (\ref{Eq:matrixForm}) as the difference equation, characterizing a linear system with $\underline{X}$ and $\underline{N}(t)$'s as its inputs, and $\underline{Z}(t)$ its output, and using the results in \cite{kailath1980linear}, $\{Z(t)\}_i$'s are given by:
\begin{equation}\label{Eq:measForm1}
\underline{Z}(t)= \Psi(t) \cdot \underline{X}+\underline{N}_{\rm{eff}}(t),
\end{equation}
where the \textit{marginal measurement} matrix, $\Psi(t)$, and the \textit{marginal effective noise} vector, $\underline{N}_{\rm{eff}}(t)$, are calculated as follows:
\begin{eqnarray}
\Psi(t)&=&B \cdot \sum_{t'=2}^{t} \prod_{t''=t}^{t'+1} F(t'') ~A(t'), \label{Eq:DefPsi} \\
\underline{N}_{\rm{eff}}(t)&=&B \cdot \sum_{t'=2}^{t} \prod_{t''=t}^{t'+1} F(t'') ~ \underline{N}(t'). \label{Eq:DefNeff}
\end{eqnarray}
In Eqs.~\ref{Eq:DefPsi},\ref{Eq:DefNeff}, the matrix multiplication is defined as:
\begin{equation}
\prod_{t''=t}^{t'+1} F(t'')=F(t) \cdots F(t'+1).
\end{equation}

By storing $\underline{Z}(t)$'s, at the decoder, we build up the \textit{total measurements vector}, $\underline{Z}_{\rm{tot}}(t)$, as follows:
\begin{equation}\label{Eq:totMeasEq}
\underline{Z}_{\rm{tot}}(t)=\left[ {\begin{array}{*{20}c}
	\underline{Z}(2) \\	
   \vdots   \\
   \underline{Z}(t)   \\
 \end{array} } \right]_{m \times 1},
\end{equation}
where $m=(t-1)|In(v_0)|$.
Therefore, the following can be established:
\begin{equation}\label{Eq:measForm2}
\underline{Z}_{\rm{tot}}(t)=\Psi_{\rm{tot}}(t) \cdot \underline{X}+\underline{N}_{\rm{eff,tot}}(t),
\end{equation}
where the $m \times n$ \textit{total measurement matrix}, $\Psi_{\rm{tot}}(t)$, and the \textit{total effective noise} vector, $\underline{N}_{\rm{eff,tot}}(t)$, are the concatenation result of marginal measurement matrices, $\Psi(t)$'s, and marginal effective noise vectors, $\underline{N}_{\rm{eff}}(t)$.
Because of our assumption to start transmission from $t=1$, $\{\underline{Z}(1)\}_i$'s are not useful for decoding, and therefore:
\begin{equation}\label{Eq:defPsiTot}
\Psi_{\rm{tot}}(t)=\left[ {\begin{array}{*{20}c}
	\Psi(2) \\	
   \vdots   \\
   \Psi(t)   \\
 \end{array} } \right], 
 \end{equation}
 \begin{equation}
\underline{N}_{\rm{eff,tot}}(t)=\left[ {\begin{array}{*{20}c}
	\underline{N}_{\rm{eff}}(2) \\	
   \vdots   \\
	\underline{N}_{\rm{eff}}(t)   \\
 \end{array} } \right].
\end{equation}

In the conventional linear network coding, the number of total measurements, $m$, is at least equal to the number of data, $n$. 
More precisely, the total measurement matrix is of full column rank, which makes us able to uniquely find a solution.\footnote{beyond the fact that there should not be any uncertainty involved as a result of noise.} 
In this paper, we are interested to investigate the feasibility of robust recovery of $\underline{X}$, when fewer number of measurements are received at the decoder than the number of messages; \textit{i.e.} $m < n$.

Considering the characteristic equation of (\ref{Eq:measForm2}), describing the QNC scenario, we can treat as a compressed sensing measurement equation.
This gives us an opportunity to apply the results in the literature of compressed sensing and sparse recovery \cite{CS,candes2007sparsity} to our QNC scenario with near-sparse messages.
However, one needs to examine the required conditions which guarantee sparse recovery in the proposed QNC scenario.
In the following, we discuss theoretical and practical feasibility of robust recovery with a compressed sensing perspective.

\section{Restricted Isometry Property}
\label{sec:RIP}

One of the advantageous of compressed sensing is to tackle a non-adaptive design for sensing of sparse signals, where the support (location of non-zero elements) is not known at the encoding side.
To pay back this non-adaptive characteristic, we may need more measurements than the exact number of non-zero elements.
Fortunately, if appropriate types of linear measurements are chosen, we can keep the required number of measurements much less than the number of messages; that is: $m \ll n$.

One of the properties that is widely used to characterize appropriate measurement matrices in the compressed sensing literature, is the \textit{Restricted Isometry Property} (RIP) \cite{candes}.
Roughly speaking, it provides a measure of norm conservation while the dimensionality is reduced \cite{baraniuk2007compressive}.
An $m \times n$ matrix $\Theta_{\rm{tot}}(t)$ is said to satisfy RIP of order $k$ with constant $\delta_{k}$, if for all $k$-sparse vectors $\underline{s}_k \in \mathbb{R}^n$, we have:
\begin{equation}
1-\delta_k \leq \frac{\vectornorm{\Theta_{\rm{tot}}(t) \underline{s}_k}^2}{\vectornorm{\underline{s}_k}^2}	\leq 1+\delta_k.
\end{equation}

\begin{remark}\label{remark:Gaussian}
Random matrices with identically and independently distributed (i.i.d) zero mean Gaussian entries are appropriate measurement matrices for compressed sensing.
Explicitly, an $m \times n$ i.i.d Gaussian random matrix, denoted $\mathbf{G}$, with entries of variance $\frac{1}{m}$, satisfies RIP of order $k$ and constant $\delta_k$, with a probability exceeding
\begin{equation}\label{Eq:Gens1}
1-e^{-\kappa_1 m},
\end{equation}
(called \emph{overwhelming} probability) if
\begin{equation}\label{Eq:Gens2}
m > \kappa_2 k \log(\frac{n}{k}).
\end{equation}
In (\ref{Eq:Gens1}), (\ref{Eq:Gens2}), $\kappa_1$ and $\kappa_2$ only depend on the value of $\delta_k$ (theorem~5.2 in \cite{simpleProof}).
\end{remark}

In \cite{naba1,naba2}, we proposed a design for local network coding coefficients, $\beta_{e,e'}(t)$'s and $\alpha_{e,v}(t)$'s, which results in an appropriate total measurement matrix, $\Psi_{\rm{tot}}(t)$, in the compressed sensing framework.
\begin{theorem}\label{th:DesignNCodes} (Theorem 3.1 in \cite{naba2})
Consider a quantized network coding scenario, in which the network coding coefficients, $\alpha_{e,v}(t)$ and $\beta_{e,e'}(t)$, are such that:
\begin{itemize}
\item $\alpha_{e,v}(t)=0,~\forall t>2,$
\item $\alpha_{e,v}(2)$'s are independent zero mean Gaussian random variables,
\item $\beta_{e,e'}(t)$'s are deterministic.
\end{itemize}
For such a scenario, the entries of the resulting $\Psi_{\rm{tot}}(t)$ are zero mean Gaussian random variables. 
Further, the entries of different columns of $\Psi_{\rm{tot}}(t)$, \textit{i.e.} $\{\Psi_{\rm{tot}}(t)\}_{i,v}$ and $\{\Psi_{\rm{tot}}(t)\}_{i',v'}$, where $v \neq v',$ are independent. $\blacksquare$
\end{theorem}

It is also numerically shown in \cite{naba2} that locally orthonormal set of $\beta_{e,e'}(t)$'s is a better choice than non-orthonormal sets; that is for all $e,e' \in Out(v)$, we have:
\begin{eqnarray}
\sum_{e'' \in In(v)} \beta_{e,e''}(t) \cdot \beta_{e',e''}(t) & = & 0,~e \neq e', \\
\sum_{e'' \in In(v)} \beta_{e,e''}^2(t)  & = & 1. 
\end{eqnarray} 

In \cite{naba2}, we established the relation between the satisfaction of RIP and the tail probability
\begin{eqnarray}
\textbf{p}_{\rm{tail}}(\Psi_{\rm{tot}}(t),\varepsilon )& = & \max_{\underline{x}',} \textbf{P}\Big( \Big | \vectornorm{\Psi_{\rm{tot}}(t) \underline{x}'}^2-1 \Big | > \varepsilon \Big), \nonumber \\
& & \mbox{subject to:}~ \vectornorm{\underline{x}'}=1 \label{Eq:tailProb}
\end{eqnarray}
by proving the following theorem.
\begin{theorem}\label{theorem:RIP1} (Theorem 4.1 in \cite{naba2})
Consider $\Psi_{\rm{tot}}(t)$ with the tail probability, as defined in (\ref{Eq:tailProb}), and an orthonormal transform matrix $\phi$.
Then, $\Theta_{\rm{tot}}(t)=\Psi_{\rm{tot}}(t) \cdot \phi$ satisfies RIP of order $k$ and constant $\delta_{k}$, with a probability exceeding,
\begin{equation}\label{Eq:lowerRIPbound1}
1- \left(
\begin{array}{c}
n\\
k
\end{array}
\right) (\frac{42}{\delta_k})^k ~ \textbf{p}_{tail}({\Psi_{\rm{tot}}(t)},\varepsilon=\frac{\delta_k}{\sqrt{2}}). ~\blacksquare
\end{equation}
\end{theorem}

By using theorem~\ref{theorem:RIP1}, we can analyze the behavior of $\Psi_{\rm{tot}}(t)$, resulting from the proposed local network coding coefficients in theorem~\ref{th:DesignNCodes}.
Specifically, we try to see if we can obtain the same tail probability as a Gaussian ensemble, with the same order of measurements.
Unfortunately, the complicated relation of local network coding coefficients and network parameters with the resulting $\Psi_{\rm{tot}}(t)$ (see Eqs.~\ref{Eq:defF}, \ref{Eq:defineAt}, \ref{Eq:DefPsi}, \ref{Eq:defPsiTot}) makes it difficult to derive a simple mathematical form for the tail probability, $\textbf{p}_{\rm{tail}}(\Psi_{\rm{tot}}(t),\varepsilon )$, and have a nice mathematical conclusion about the required number of measurements.

\begin{figure*}[ht]
\centering
\subfigure[$1100$ edges]{
\resizebox{!}{.41\textwidth}{
\includegraphics{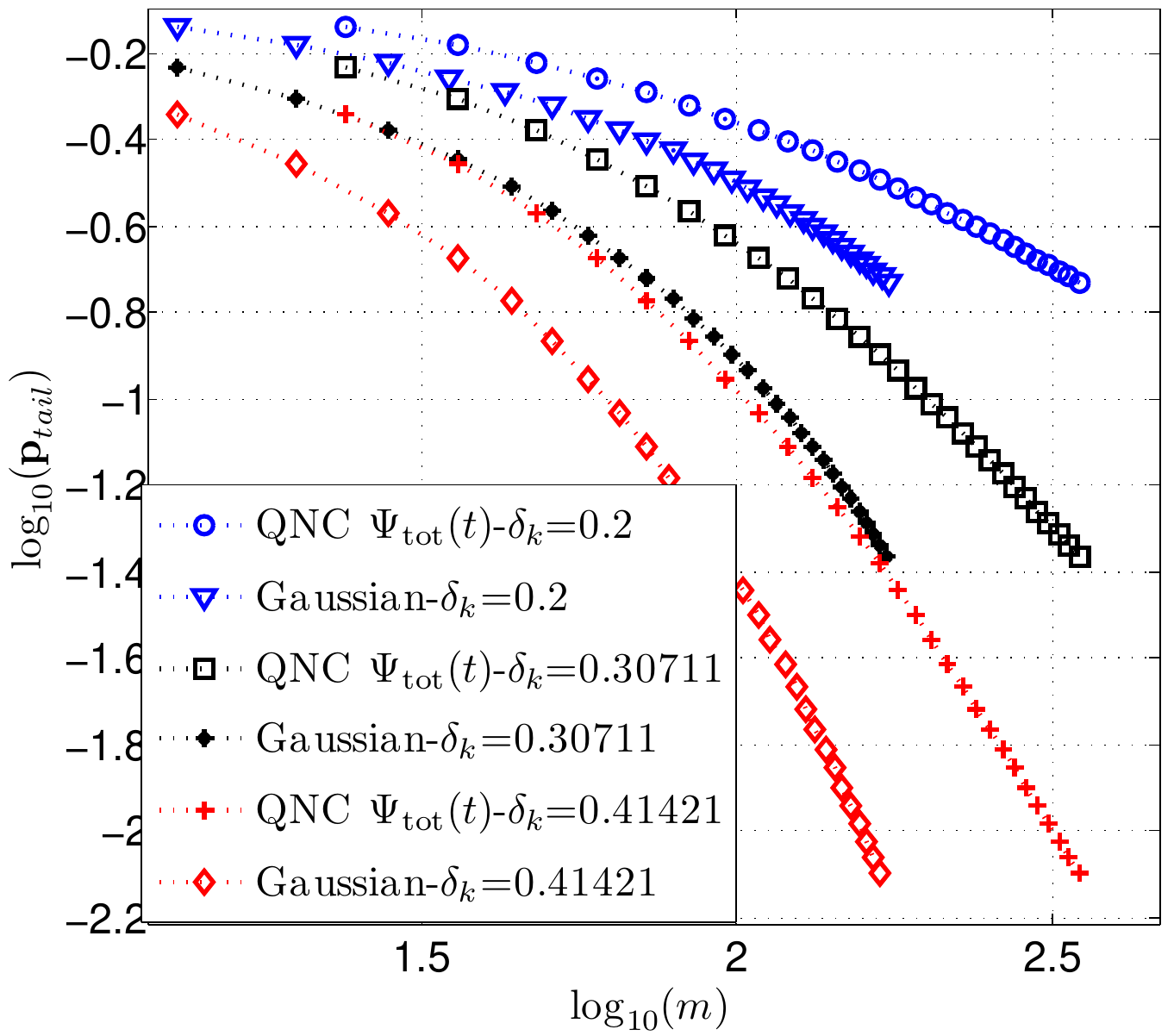}}
\label{fig:subfig1100}
} 
\subfigure[$1400$ edges]{
\resizebox{!}{.41\textwidth}{
\includegraphics{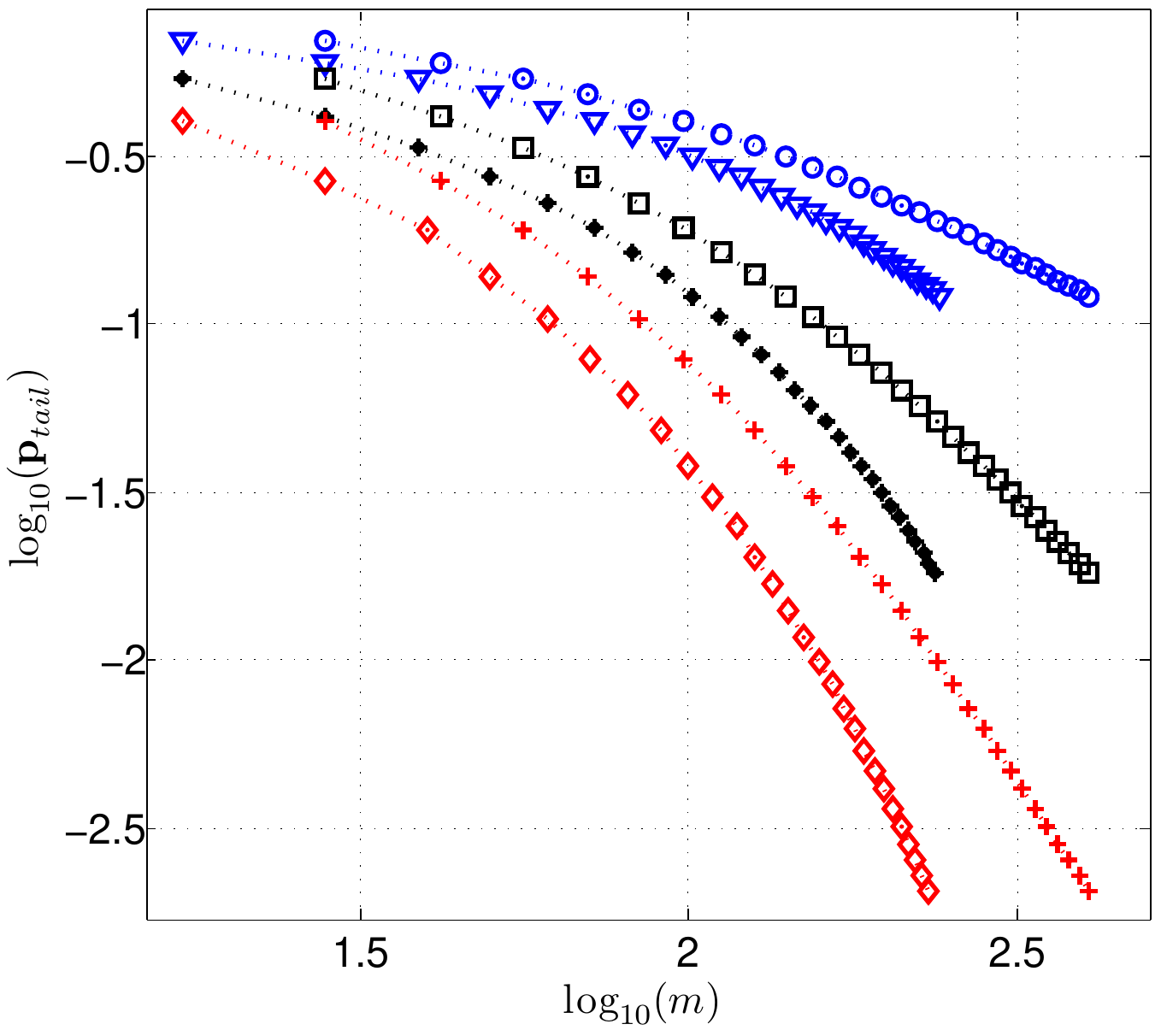}}
\label{fig:subfig1400}
} 
\subfigure[$1800$ edges]{
\resizebox{!}{.41\textwidth}{
\includegraphics{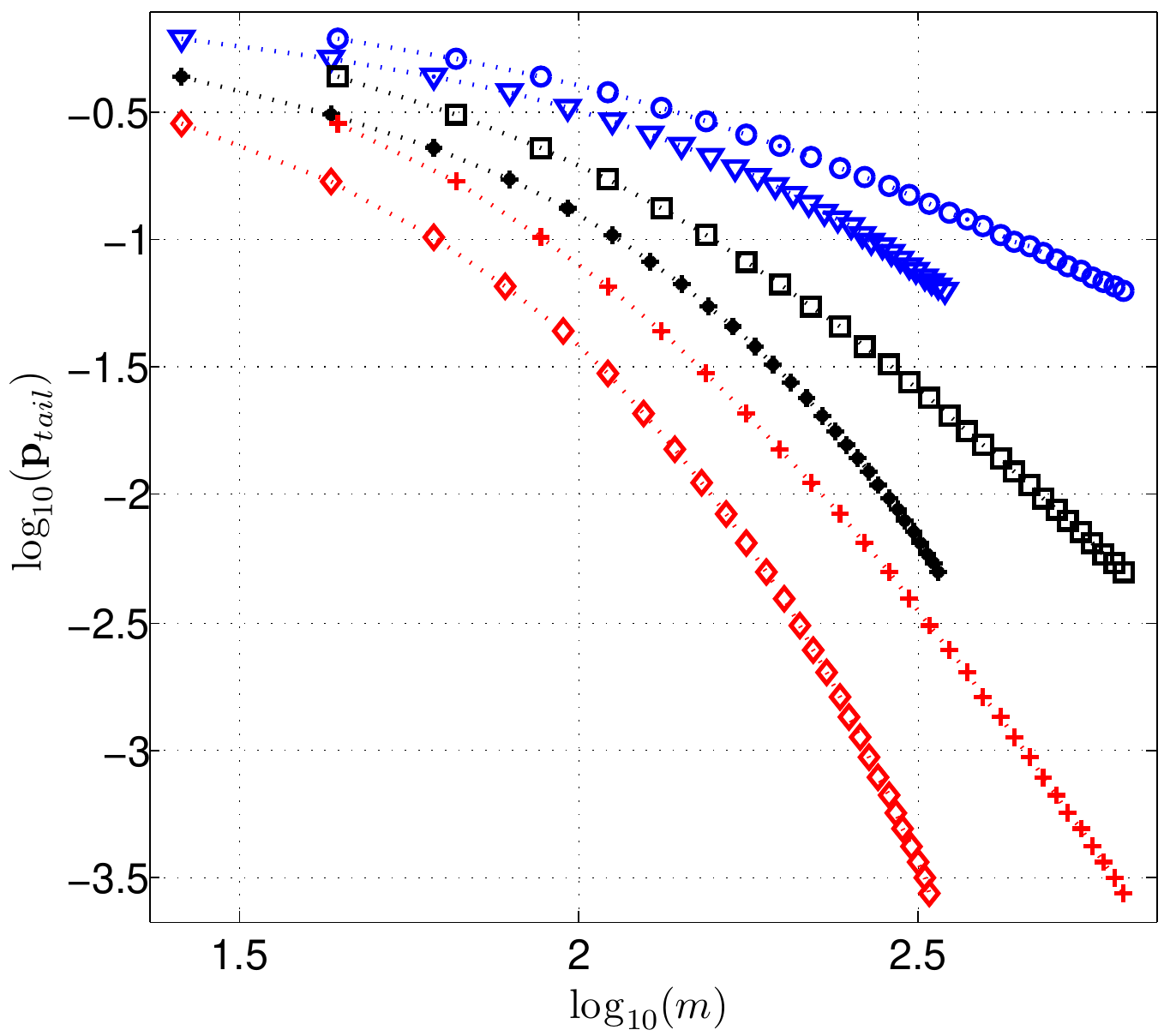}}
\label{fig:subfig1800}
} 
\caption{Logarithmic tail probability versus logarithmic ratio of minimum required number of measurements in
our QNC scenario and i.i.d. Gaussian measurement matrices, for $n=100$ nodes, different RIP constants, and different number of edges
\label{fig:tailProbs}.
}
\end{figure*}

In Fig.~\ref{fig:tailProbs}, we present the numerical values of tail probabilities (defined in Eq.~\ref{Eq:tailProb}) for the resulting $\Psi_{\rm{tot}}(t)$, $\textbf{p}_{\rm{tail}}(\Psi_{\rm{tot}}(t),\varepsilon )$, using the proposed local network coding coefficients in theorem~\ref{th:DesignNCodes}.
These tail probabilities are compared with those of i.i.d Gaussian matrices, $\mathbf{G}$, versus the number of measurements, $m$, in each case.\footnote{Detailed version of our calculations for the tail probability of $\Psi_{\rm{tot}}(t)$ can be found in \cite{naba2}.}
\begin{remark}\label{remark:RIP}
Our numerical evaluations in Fig.~\ref{fig:tailProbs} show that for the same value of tail probability, there is a QNC resulting measurement matrix, $\Psi_{\rm{tot}}(t)$, and an i.i.d Gaussian matrix, $\mathbf{G}$, which have the same order of measurements, $m$.
Furthermore, using theorem~\ref{theorem:RIP1}, we can say the the resulting $\Psi_{\rm{tot}}(t)$ has a similar behavior as Gaussian matrices, in terms of RIP satisfaction.
\end{remark}

In the following section, we use the aforementioned conclusion for the resulting $\Psi_{\rm{tot}}(t)$ to derive a performance bound for QNC scenario.

\section{Decoding using Sparse Recovery}
\label{sec:L1Dec}

Sparse recovery for exactly sparse data can be done by using linear programming \cite{LinProg}, where NP-hard $\ell_0$ minimization is replaced by $\ell_1$ minimization. 
Fortunately, this alteration of cost function does not affect the optimality in recovery of exactly sparse vectors from noiseless measurements \cite{LinProg,candes2007sparsity}.
However, when dealing with noisy measurements, $\ell_1$-min recovery does not necessarily offer an optimal solution.
There is still a lot of work being done to develop practical and near-optimal recovery algorithms for noisy cases.
In the following, we discuss $\ell_1$-min recovery for QNC scenario and establish theoretical bounds on its recovery error.

Motivated by the work in \cite{CS,candes}, the compressed sensing based decoder for QNC scenario solves the following convex optimization:
\begin{eqnarray}
\underline{\hat{X}}(t)&=&\phi \cdot \arg\min_{\underline{S}'} \vectornormOne{\underline{S}'}, \label{Eq:L1minDecoder} \\
 && \text{subject to:}~\vectornorm{\underline{Z}_{\rm{tot}}(t)-\Psi_{\rm{tot}}(t)~ \phi ~ \underline{S}'} \leq \epsilon_{rec}(t) \nonumber
\end{eqnarray}
which can be solved by using linear programming \cite{LinProg}.
In the following, we present our results on the recovery error using $\ell_1$-min decoding of Eq.~\ref{Eq:L1minDecoder}.
\begin{theorem}\label{th:RecErrBnd}
Consider a QNC scenario where, for all $v \in \mathcal{V}$, the network coding coefficients satisfy the following normalization condition:
\begin{equation}\label{Eq:Normalization}
\sum_{e' \in In(v)} |\beta_{e,e'}(t)| + |\alpha_{e,v}(t)| \leq 1,~\forall e \in Out(v),~\forall t.
\end{equation}
In such scenario, we assume that the resulting $\Theta_{\rm{tot}}(t)=\Psi_{\rm{tot}}(t) \phi$ satisfies RIP of order $2k$ with constant $\delta_{2k}$, where $\delta_{2k} < \sqrt{2}-1$.
Moreover, the messages, $X_v$'s, are supposed to be bounded between $-q_{max}$ and $+q_{max}$, and the edge quantizers, $\textbf{Q}_e(\centerdot)$'s, are uniform with the step size 
\begin{equation}
\Delta_e=\frac{2q_{max}}{\lfloor 2^{L C_e} \rfloor}.
\end{equation}
Now, for the $\ell_1$-min decoding of (\ref{Eq:L1minDecoder}) where $\epsilon^2_{rec}(t)$ is as defined in Eq.~\ref{Eq:defEpsRec},
\begin{figure*}[t]
\begin{equation}\label{Eq:defEpsRec}
\epsilon^2_{rec}(t)= \frac{1}{4}  \sum_{t'=2}^{t} \Big ( \sum_{t''=1}^{t'-1} \underline{\Delta}_Q^{T}~ {\Big | \prod_{t'}^{t'''=t''+2}{F(t''')} \Big |}^{T}  
 \cdot B^T B \cdot \sum_{t''=1}^{t'-1} \Big | \prod_{t'}^{t'''=t''+2}{F(t''')} \Big | ~ \underline{\Delta}_Q \Big )
\end{equation}
\end{figure*}
and $\underline{\Delta}_Q = [\Delta_{e}:e \in \mathcal{E}]$, we have:
\begin{equation}\label{Eq:recErrBnd1}
\vectornorm{\underline{\hat{X}}(t)-\underline{X}} \leq c_1 \epsilon_{rec}+\frac{c_2}{\sqrt{2}} \epsilon_{k}.
\end{equation}
In the inequality of (\ref{Eq:recErrBnd1}), $c_1$ and $c_2$, are constants, defined as follows:
\begin{eqnarray}
c_1 &=& 4 \frac{\sqrt{1+\delta_{2k}}}{1-(1+\sqrt{2})\delta_{2k}}, \\
c_2 &=& 2 \frac{1-(1-\sqrt{2})\delta_{2k}}{1-(1+\sqrt{2})\delta_{2k}}.
\end{eqnarray}
\end{theorem}

\begin{proof}
Since the network is lossless and network coding coefficients satisfy the condition of Eq.~\ref{Eq:Normalization}, and $|X_v| \leq {q}_{max},~\forall~v$, the only associated measurement noise is resulting from the quantization noise at the edges.
Moreover, for each $e \in \mathcal{E}$, we have:
\begin{equation}
-\frac{\Delta_{e}}{2} \leq N_{e}(t) \leq +\frac{\Delta_{e}}{2},
\end{equation}
since the quantizers are uniform.
Equivalently, the absolute value vector of $\underline{N}(t)$, represented by $\Big|\underline{N}(t)\Big|$, is such that:
\begin{equation} \label{Eq:absUpB0}
\Big| \underline{N}(t)\Big| \leq \frac{1}{2} \underline{\Delta}_Q.
\end{equation}
Since $B$ is a one-to-one mapping matrix with positive entries, the effective noise vector, $\underline{N}_{\rm{eff}}(t)$, can be upper bounded as follows:
\begin{eqnarray}
\Big | \underline{N}_{\rm{eff}}(t) \Big| &=& \Big | B~\sum_{t'=1}^{t-1} \prod_{t}^{t''=t'+2}{F(t'')}~ \underline{N}(t'+1) \Big| \label{Eq:absUpB1}  \\
& = & B \cdot \Big | \sum_{t'=1}^{t-1} \prod_{t}^{t''=t'+2}{F(t'')}~ \underline{N}(t'+1) \Big|. \label{Eq:absUpB2}  
\end{eqnarray}
Therefore, using (\ref{Eq:absUpB0}), we have:
\begin{eqnarray}
\Big | \underline{N}_{\rm{eff}}(t) \Big| & \leq & B \cdot  \sum_{t'=1}^{t-1} \Big | \prod_{t}^{t''=t'+2}{F(t'')} ~ \underline{N}(t'+1) \Big | \label{Eq:absUpB3}   \\
& \leq & B \cdot  \sum_{t'=1}^{t-1} \Big | \prod_{t}^{t''=t'+2}{F(t'')} \Big | \cdot  \Big | \underline{N}(t'+1) \Big | \label{Eq:absUpB4}   \\
& \leq & B \cdot  \sum_{t'=1}^{t-1} \Big | \prod_{t}^{t''=t'+2}{F(t'')} \Big | \cdot \frac{1}{2} \underline{\Delta}_Q. \label{Eq:absUpB5} 
\end{eqnarray}
Finally, by using inequality of (\ref{Eq:absUpB5}), 
\begin{equation}
\vectornorm{\underline{N}_{\rm{eff,tot}}(t)}^2 =  \sum_{t'=2}^{t} \vectornorm{\underline{N}_{\rm{eff}}(t')}^2,
\end{equation}
and,
\begin{equation}
\vectornorm{\underline{N}_{\rm{eff}}(t')}^2={\Big | \underline{N}_{\rm{eff}}(t) \Big |}^{T} \cdot \Big | \underline{N}_{\rm{eff}}(t) \Big |,
\end{equation}
we can show that:
\begin{equation}
\vectornorm{\underline{N}_{\rm{eff,tot}}(t)}^2 \leq \epsilon^2_{\rm{rec}}(t),
\end{equation}
such that $\epsilon_{\rm{rec}}(t)$ is as defined in Eq.~\ref{Eq:defEpsRec}.
Now, by applying theorem~4.2 in \cite{CSbook}, and using the definition of $\epsilon_k$ in (\ref{Eq:defEpsK}), we can finish the proof of our theorem.
\end{proof}

According to the preceding theorem, the upper bound, $c_1 \epsilon_{\rm{rec}}$, is decreased when the quantization steps, $\Delta_{e}$'s, are decreased, too. And since 
$\Delta_e={2q_{max}}/{\lfloor 2^{L C_e} \rfloor}$, a smaller upper bound on the $\ell_2$-norm of recovery error is forced by increasing the block length, $L$.
Although this can be done practically, it will simultaneously increase the point to point transmission delays in the network, which may not be desirable.
Introducing a trade-off on the choice of block length, one has to find its appropriate value for a specific quality of service (\textit{i.e.} recovery error).



Based on theorem~\ref{th:RecErrBnd}, if the resulting $\Psi_{\rm{tot}}(t)$ satisfies RIP of appropriate order with a high probability, then the robust recovery can be guaranteed.
On the other hand, using remarks~\ref{remark:Gaussian},\ref{remark:RIP}, we can say that the resulting $\Psi_{\rm{tot}}(t)$ satisfies RIP with a high probability, while the number of measurements, $m$, has a smaller order than the number of messages, $n$.
Therefore, putting all these numerical and theoretical results together, it is true to say that QNC can result in bounded error recovery (\ref{Eq:recErrBnd1}) with a smaller order of measurements (received packets at the decoder) than that of messages.
This saving in the required number of received packets can be interpreted as an \textit{embedded distributed compression}, achieved by quantized network coding at the nodes.

\section{Simulation Results}
\label{sec:SimRes}

In this section, we evaluate the performance of quantized network coding, by using different numerical simulations.
We are interested to find out the compression achievements, resulting from QNC, by obtaining the delay-distortion curves in different scenarios.

Although we were able to derive mathematical performance measures for the QNC scenario, they are not comprehensive and do not offer any guarantee on the statistical performance measures; \textit{e.g.} mean squared error.
However, deriving such statistical performance bounds requires a lot more of theoretical work on the sparse recovery, and meanwhile we can only rely on the numerical evaluations.

We initiate our numerical evaluations, by comparing the delay-quality performance of QNC and conventional routing based packet forwarding for lossy transmission of a set of correlated sources (messages).
To set up the simulations, we randomly generate random deployments of directed networks with uniformly distributed edges (making sure there is not any pair of nodes with two assigned edges). 
The edges can maintain a lossless communication of one bit per use, meaning $C_e=1$, for all $e \in \mathcal{E}$.
One of the nodes is randomly picked to be the gateway node, $v_0$, in which the messages are decoded.
To generate a realization of messages, $\underline{x}$, we first generate a $k$-sparse random vector, $\underline{s}_k$, whose components are uniformly distributed between $-\frac{1}{2}$ and $+\frac{1}{2}$.
A near-sparse vector, $\underline{s}$, is obtained by adding a zero mean uniform noise, such that $\vectornorm{\underline{s}-\underline{s}_k}$ is bounded by $\epsilon_k$.
This is followed by generation of an orthonormal random matrix, $\phi$, and calculating random messages; $\underline{x}=\phi \cdot \underline{s},$ and normalizing the range of $x_v$'s, between $-q_{max}$ and $+q_{max}$, where $q_{max}=10$.
Different values of sparsity factor, $\frac{k}{n}$, and $\epsilon_k$ are used in our simulations.
A summary of the simulation parameters is presented in Table~\ref{Table:simParams}.
\begin{table}
  \caption{The parameters of messages and the networks, used in our simulations.\label{Table:simParams}}
\begin{center}
\setlength{\extrarowheight}{1.5pt}
\begin{tabular}{|c ||c|}
\hline 
Parameter & Value(s) \\ 
\hline \hline
No of nodes, $n$ & $100$ \\ \hline 
No of edges, $|\mathcal{E}|$ & $1100,1400,1800$ \\ \hline 
Block length, $L$ & $1,\ldots,40$ \\ \hline 
Sparsity factor, $k/n$ & $0.05,0.15,0.25$ \\ \hline 
Near-sparsity factor, $\epsilon_k/\vectornormOne{\underline{s}_k}$ & $0,0.002,0.02,0.2$ \\ \hline
Range of messages & $[-10,+10]$ \\ \hline 
Average $\vectornorm{\underline{x}}$ & $41.7 \sim 32.4$ [dB] \\ \hline
\end{tabular} 
\end{center}
\end{table}

For each generated random network deployment, we perform QNC with $\ell_1$-min decoding.
Local network coding coefficients, $\alpha_{e,v}(t)$'s and $\beta_{e,e'}(t)$'s, are generated according to the conditions of theorem~\ref{th:DesignNCodes}.
The freedom degrees are limited by picking $\beta_{e,e'}(t)$'s such that they are locally orthonormal; 
The resulting coefficients are then normalized to satisfy the normalization condition of Eq.~\ref{Eq:Normalization} and prevent overflow in the linear combination of QNC.
Edge quantizers, $\textbf{Q}_{e}(\centerdot)$'s, are uniform with a range of $[-q_{max},+q_{max}]$ and $2^{L}$ intervals (since $C_e=1,~\forall e$).
This completes all the required parameters and vectors to simulate quantized network coding and obtain the received packets at the decoder node, $\underline{z}(t)$'s.\footnote{Lower case notations are used for realization of random variables.}
Random $\alpha_{e,v}(2)$'s can be generated in a pseudo-random way and therefore only the generator seed needs to be transmitted to the decoder as a header.

At the decoder, the received measurements up to $t$, $\underline{z}_{\rm{tot}}(t)$, are used to recover the original messages.
Specifically, for a realization of messages, $\underline{x}$, we define $\underline{\hat{x}}_{\rm{QNC}}(t)$, to be the recovered messages, using $\ell_1$-min decoding, according to (\ref{Eq:L1minDecoder}).
Moreover, the convex optimization, involved in (\ref{Eq:L1minDecoder}) is solved by using the open source implementation of disciplined convex programming \cite{cvx1,cvx2}.

For each deployment, we also simulated a routing based packet forwarding and compare it with the results for QNC.
To find the routes from each node to the gateway node, we find the shortest path from each node to the gateway node, using the Dijkstra algorithm \cite{dijkstra1959note}.
Further, the real valued messages, $x_v$'s, are quantized at their corresponding source nodes, by using similar uniform quantizers, as used in QNC transmission.
It is aimed to deliver all $x_v$'s to the decoder node and keep the track of delivered messages over time, $t$, in the recovered vector of messages, $\underline{\hat{x}}_{\rm{PF}}(t)$.
Moreover, if a message, $x_v$, is not delivered by time index $t$, zero is used as its recovered value: 
\begin{equation}
\{\underline{\hat{x}}_{\rm{PF}}(t)\}_{v}=0.
\end{equation}

The $\ell_2$ norm of recovery error, $\vectornorm{\underline{x}-\underline{\hat{x}}(t)}$, is used as the quality measure in our numerical comparisons.
The payback measure in our comparisons is the delivery delay, corresponding to achieve a minimum quality of service.
Explicitly, delivery delay for a transmission which has terminated at $t$ is equal to $(t-1)  L$ in both cases of QNC and packet forwarding.\footnote{In the case of packet forwarding, we do not consider the learning period, required to find the optimal routes.}
In QNC scenario, for each value of $\frac{k}{n}$, and $\epsilon_k$, we calculate the average of $\vectornorm{\underline{x}-\underline{\hat{x}}_{\rm{QNC}}(t)}$'s over different realizations of network deployments.
Since the sparsity of messages does not affect the performance of packet forwarding, we only need to present its results for different network parameters (\textit{i.e.} number of edges).

\begin{figure*}[t]
\centering
\subfigure[$1100$ edges]{
\resizebox{.43\textwidth}{!}{
\includegraphics{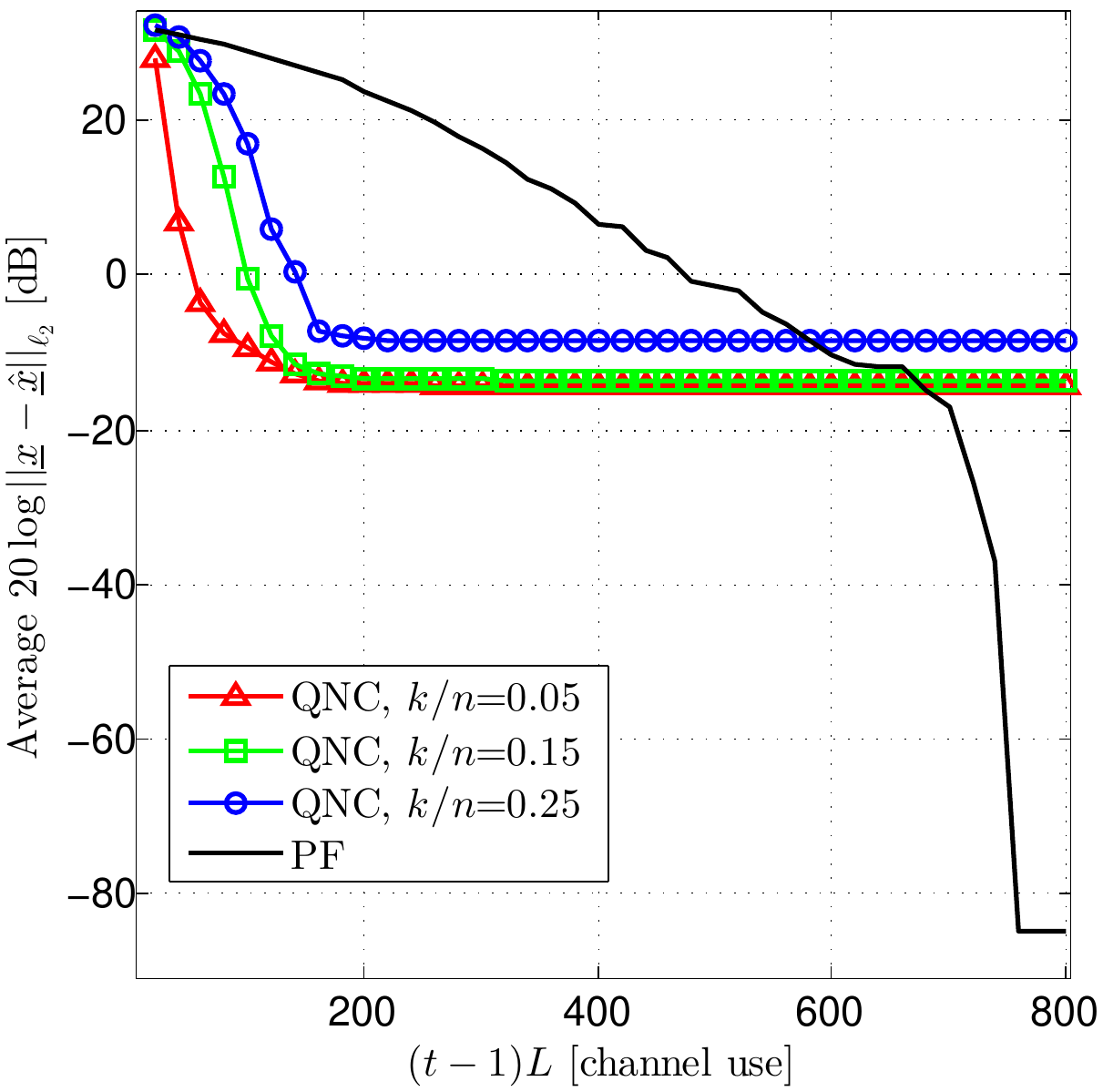}}
\label{fig:marginal1}
} \qquad
\subfigure[$1400$ edges]{
\resizebox{.43\textwidth}{!}{
\includegraphics{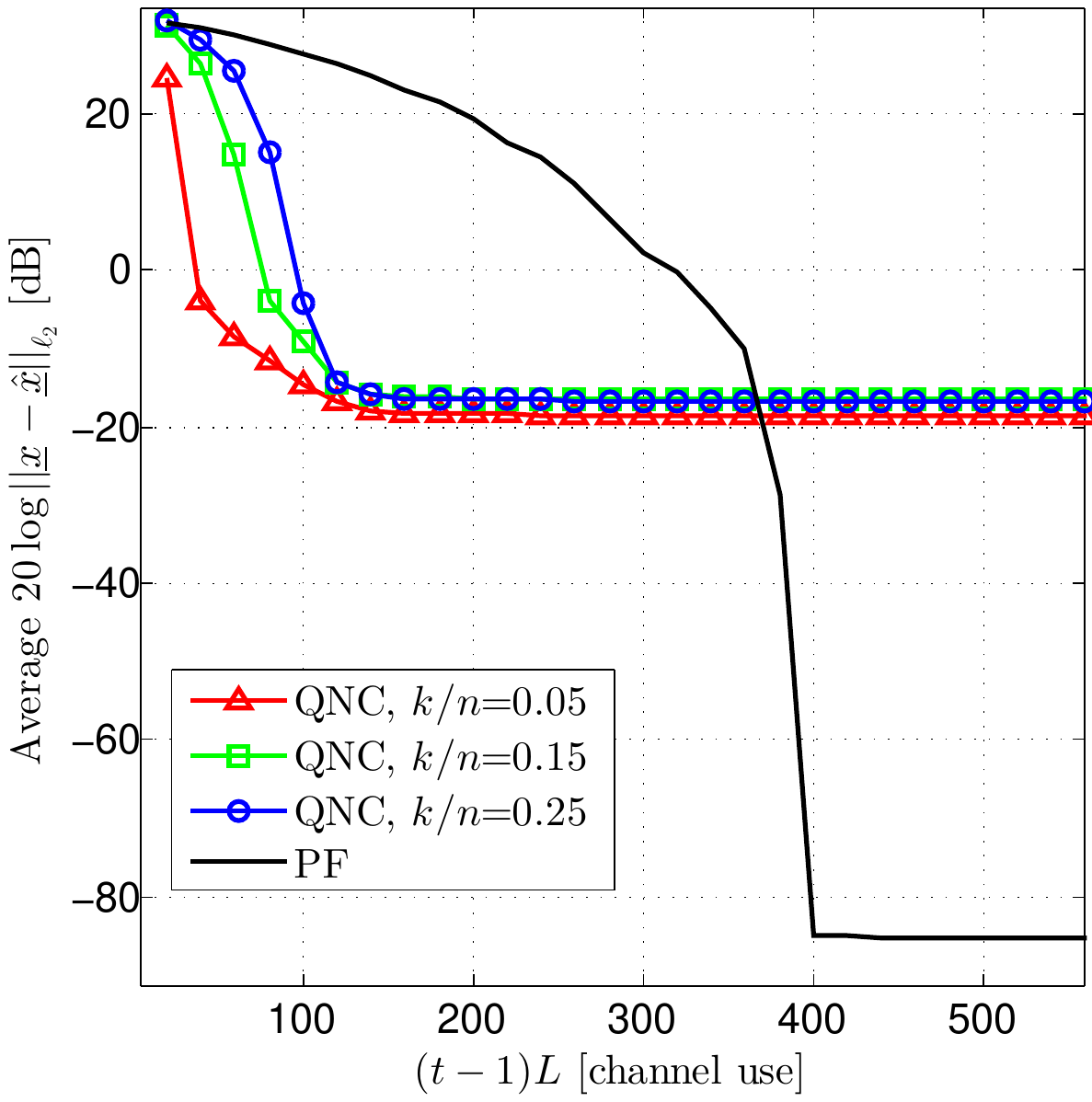}}
\label{fig:marginal2}
} 
\caption{Average $\ell_2$-norm of recovery error versus delivery delay for QNC and PF scenarios, when $L=20$, $\epsilon_k=0$ and $|\mathcal{E}|=1100,1400$\label{fig:marginalRes}.
}
\end{figure*}

For a fixed block length, $L=20$, the average of $\ell_2$ norm of recovery error versus the average delivery delay is depicted in Fig.~\ref{fig:marginalRes}.
In Figs.~\ref{fig:marginal2},\ref{fig:marginal1}, the horizontal axis represents the product $(t-1)L$, which is the delivery delay, corresponding to $L=20$, for different values of $t \geq 1$.
The vertical axis is the logarithmic average $\ell_2$ norm of recovery error, $20\log_{10} \vectornorm{\underline{x}-\underline{\hat{x}}}$, for QNC and Packet Forwarding (PF) scenarios.
Simply, Figs.~\ref{fig:marginal2},\ref{fig:marginal1} can be considered as rate-distortion curves in a lossy source coding scenario.

As it is shown in Figs.~\ref{fig:marginal2},\ref{fig:marginal1}, when using the same block length, QNC achieves significant improvement, compared to PF, for low values of delivery delay.
These low delays correspond to the initial $t$'s in the transmission, at which a small number of packets are received at the decoder, as expected.
After enough packets are received at the decoder, QNC achieves its best performance (at around $-20$ [dB]), as a result of associated quantization noises.
The best performance for packet forwarding happens after a longer period of time than for QNC.
On the other hand, the best performance of PF (around $-80$ [dB]) is higher than that of QNC, which can be explained by noise propagation in the network during QNC.
However, as it is shown in the following, QNC outperforms PF in a wide range of delay values, when the appropriate block length is adopted in each case.

\begin{figure*}[t]
\centering
\subfigure[$1100$ edges, $\epsilon_k=0$]{
\resizebox{.39\textwidth}{!}{
\includegraphics{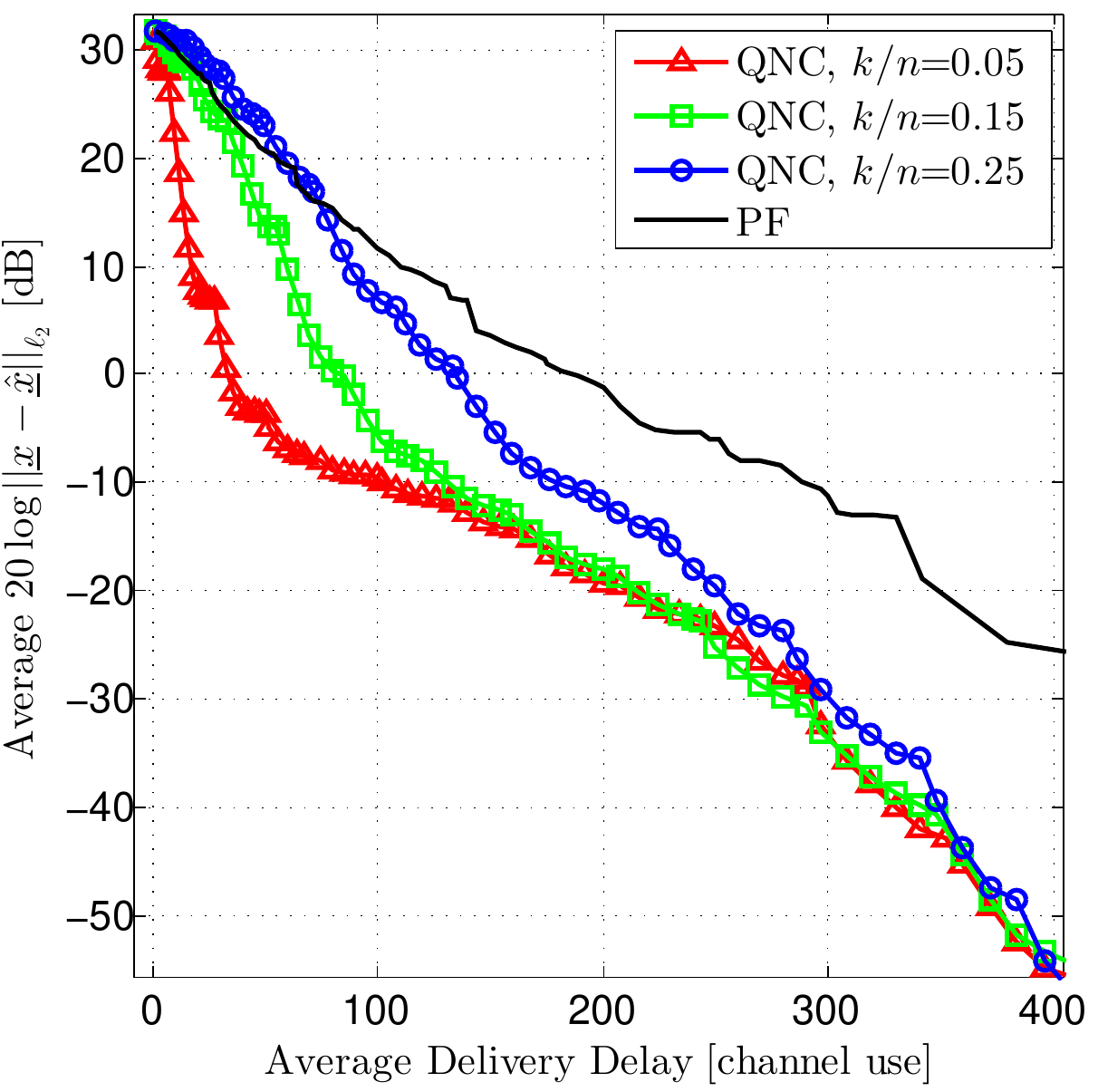}}
\label{fig:1100edgesEpsk1}
} \qquad
\subfigure[$1100$ edges, $\epsilon_k=0.2$]{
\resizebox{.39\textwidth}{!}{
\includegraphics{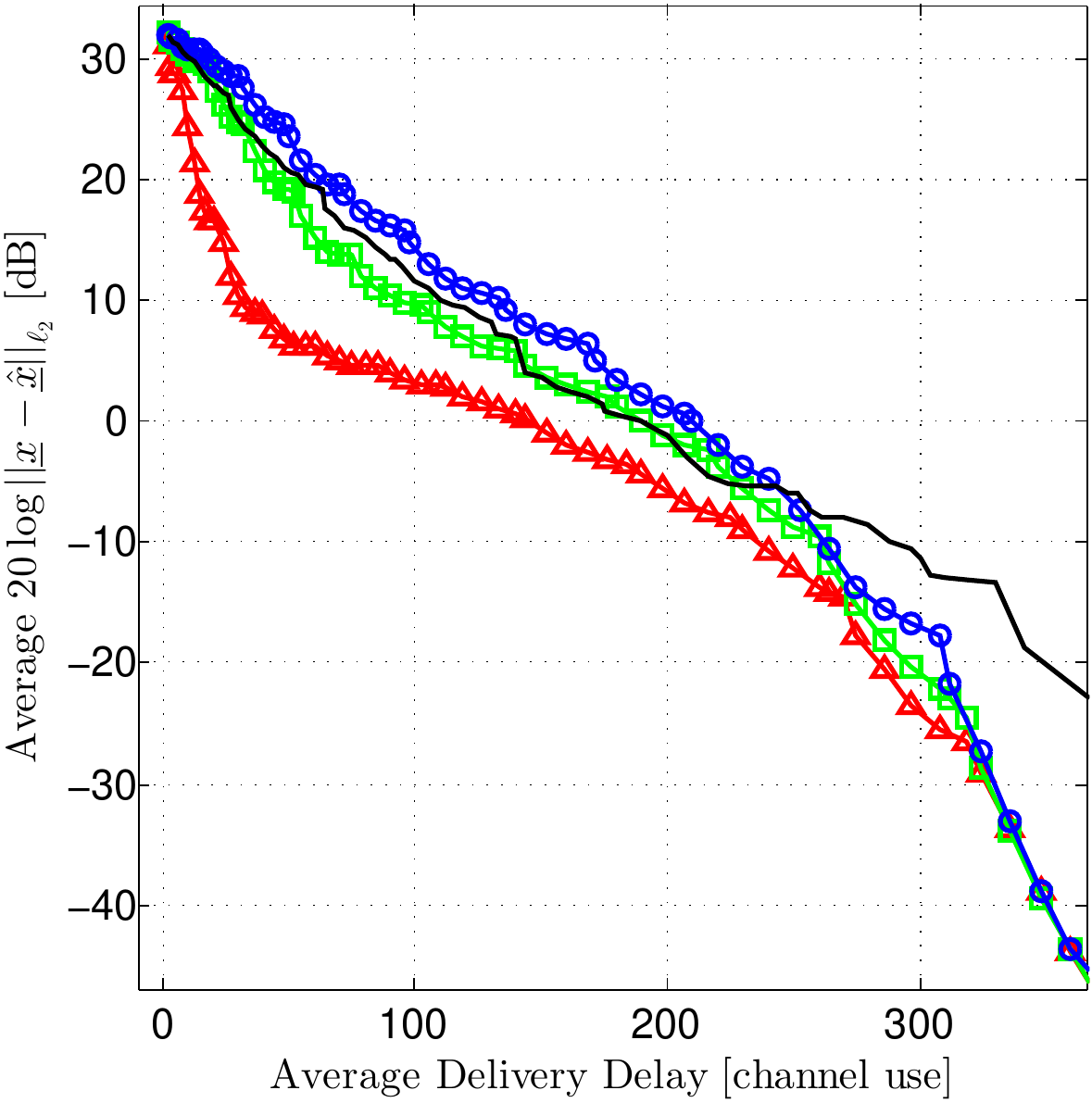}}
\label{fig:1100edgesEpsk4}
} \\
\subfigure[$1400$ edges, $\epsilon_k=0$]{
\resizebox{.39\textwidth}{!}{
\includegraphics{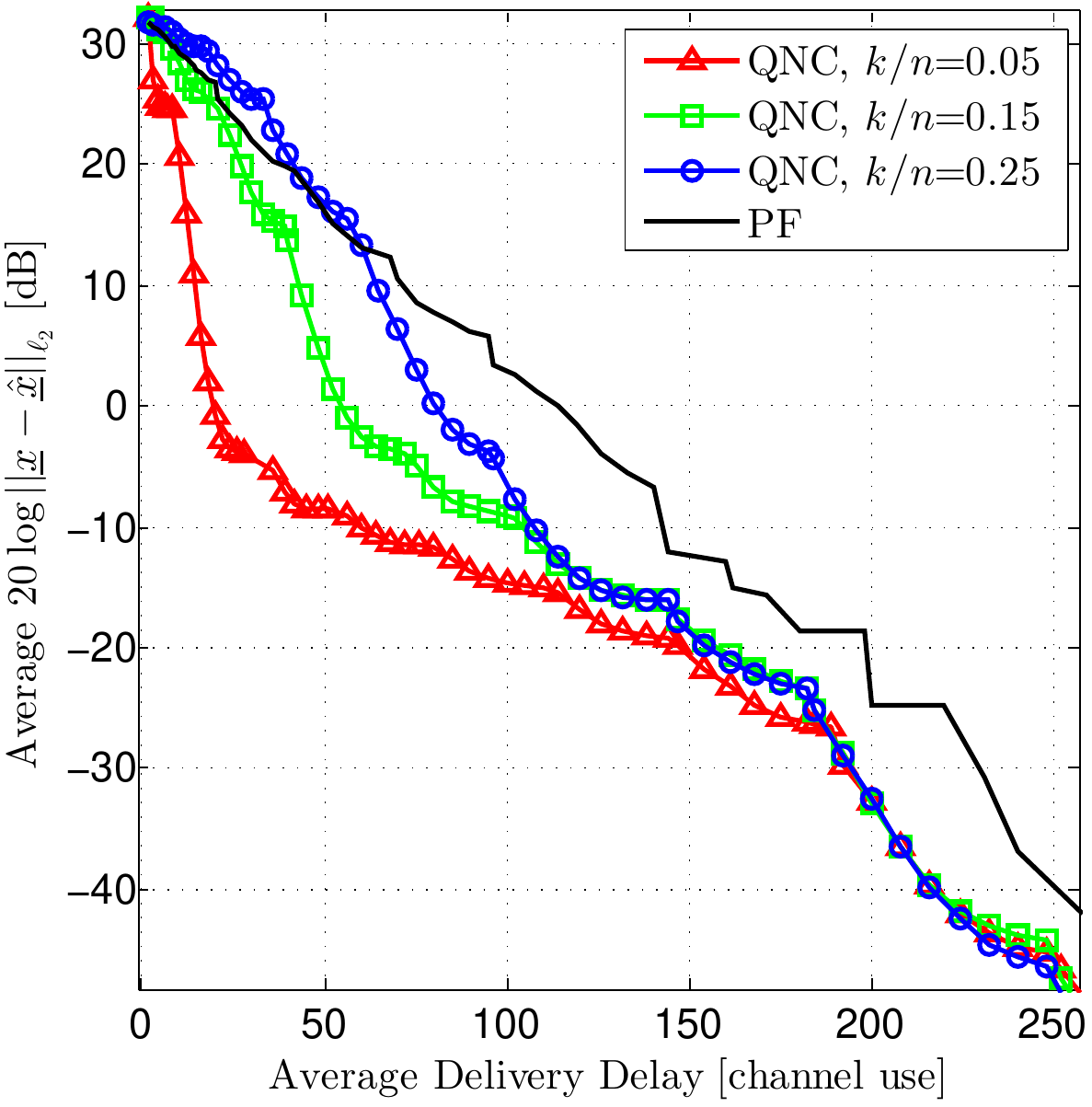}}
\label{fig:1400edgesEpsk1}
} \qquad
\subfigure[$1400$ edges, $\epsilon_k=0.2$]{
\resizebox{.39\textwidth}{!}{
\includegraphics{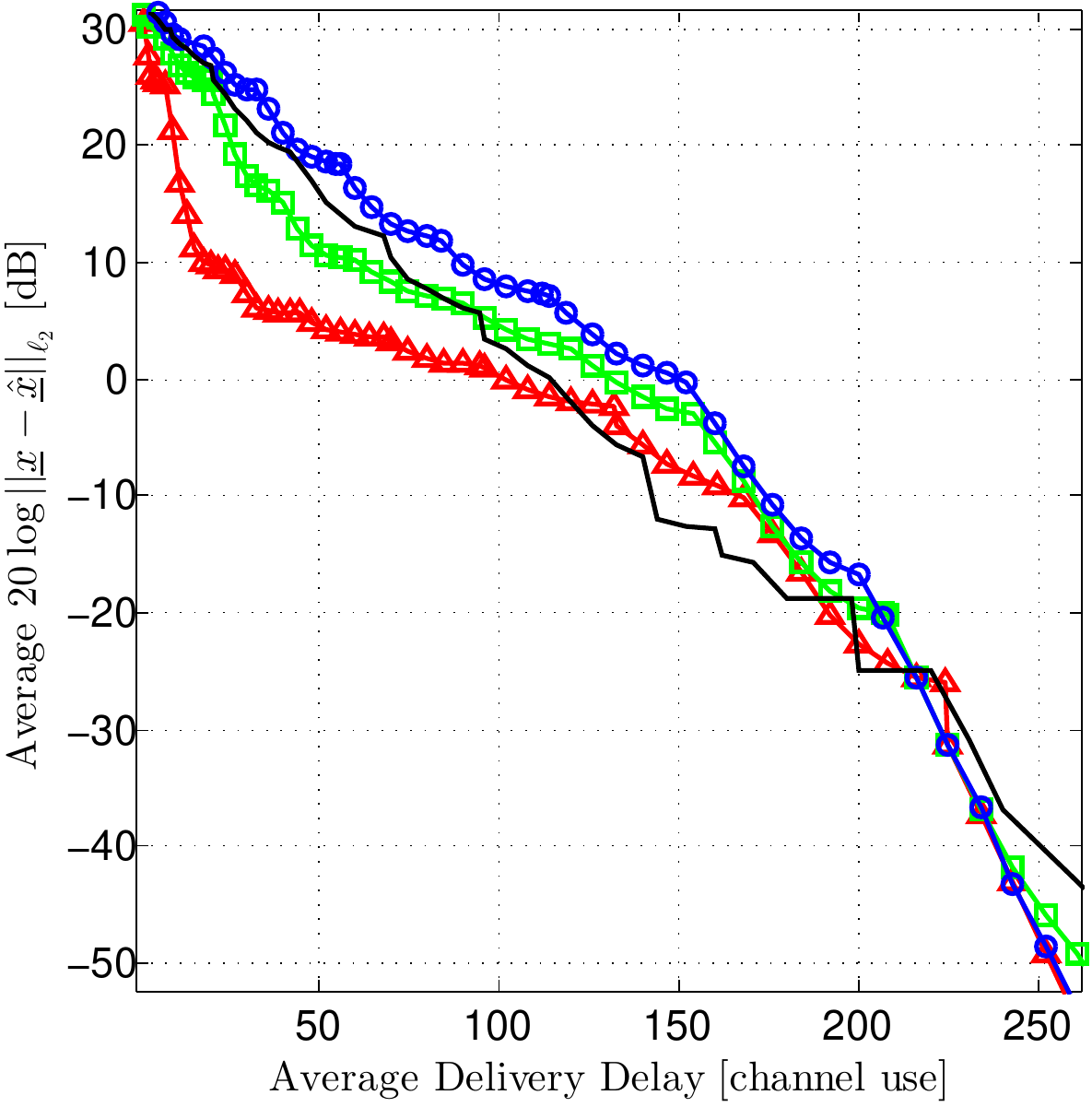}}
\label{fig:1400edgesEpsk4}
} \\
\subfigure[$1800$ edges, $\epsilon_k=0$]{
\resizebox{.39\textwidth}{!}{
\includegraphics{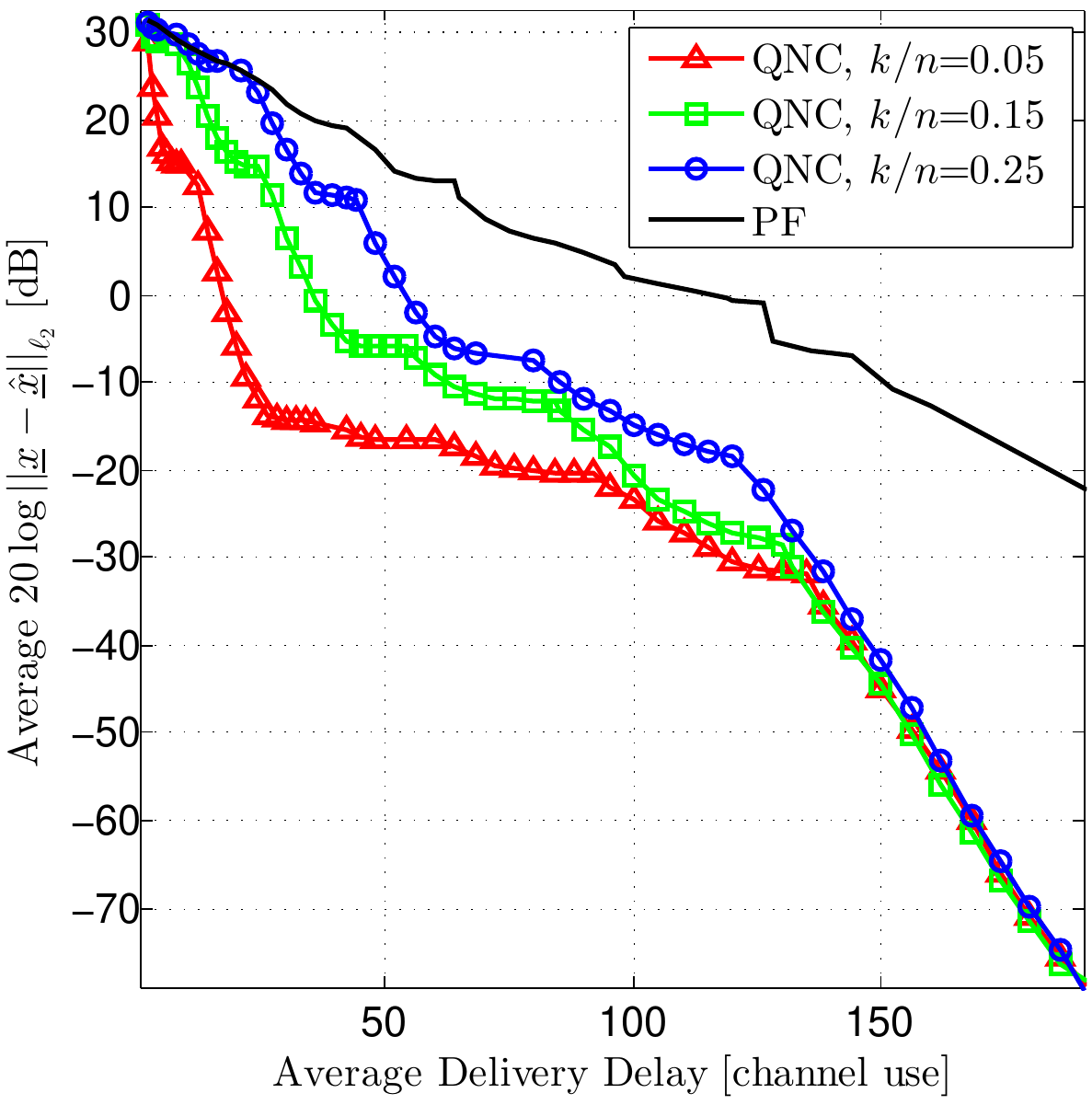}}
\label{fig:1800edgesEpsk1}
} \qquad
\subfigure[$1800$ edges, $\epsilon_k=0.2$]{
\resizebox{.39\textwidth}{!}{
\includegraphics{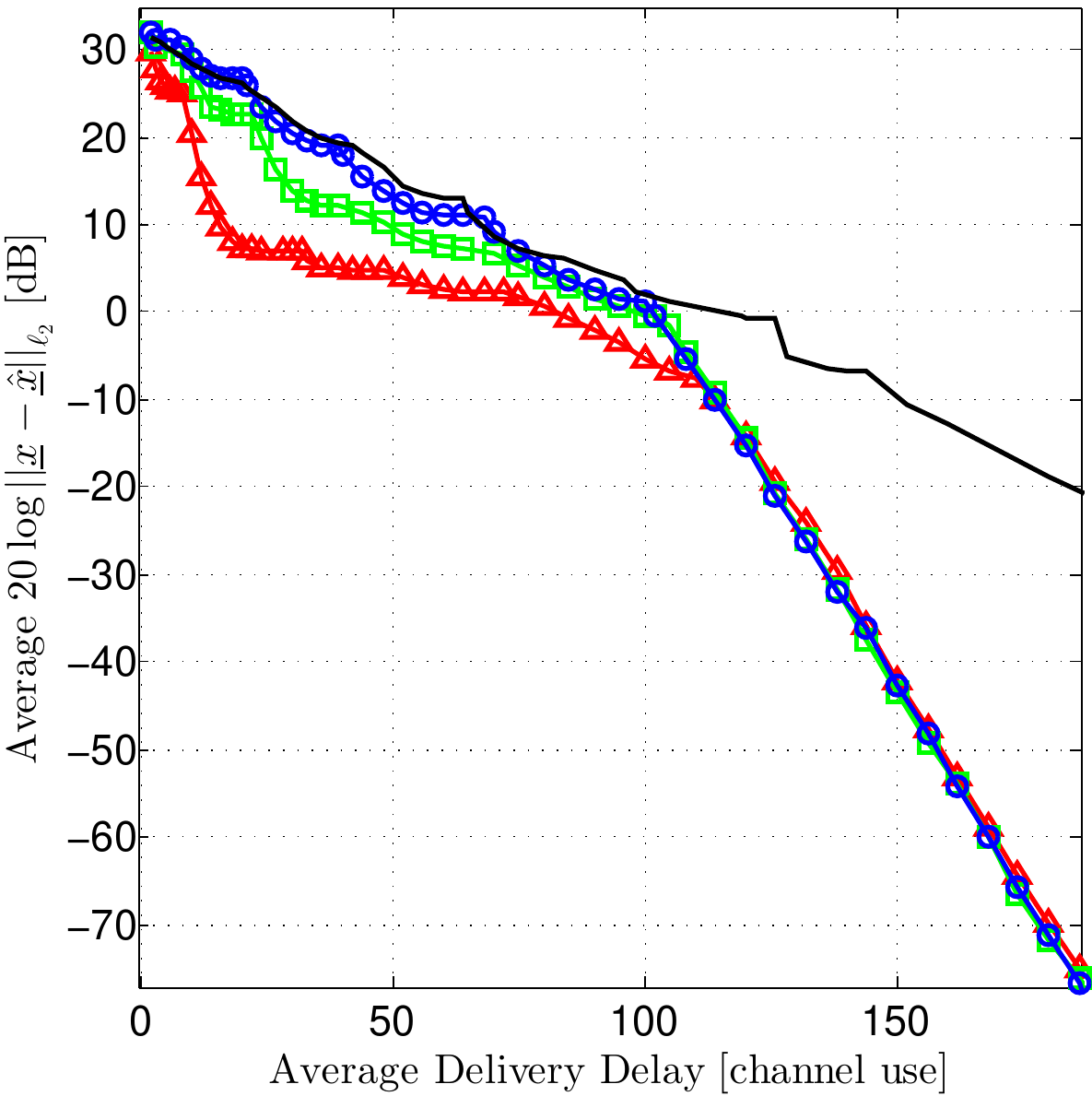}}
\label{fig:1800edgesEpsk4}
} \\
\caption{Average $\ell_2$-norm of recovery error versus average delivery delay for QNC and PF, and different values of $\epsilon_k$\label{fig:Overall}.}
\end{figure*}

After simulating QNC and PF scenarios for different block lengths and calculating the corresponding delay and recovery error norms, we find the best values of block length for each specific average $\ell_2$ norm of recovery error (as a measure of quality of service).
The resulting $L$-optimized curve for each QNC and PF scenario is depicted in Fig.~\ref{fig:Overall}.
In Figs.~\ref{fig:1100edgesEpsk1}-\ref{fig:1800edgesEpsk4}, QNC performance is compared with that of PF, for different number of edges, different sparsity factors, and near-sparsity parameters.
Generally speaking, these figures show a promising improvement over conventional packet forwarding, when QNC is adopted for transmission of near-sparse messages.
The achieved improvement is increased as the sparsity factor, $\frac{k}{n}$, is decreased, meaning a higher level of correlation between the messages.

\begin{figure*}[t]
\centering
\subfigure[$1100$ edges, $\frac{k}{n}=0.05$]{
\resizebox{.41\textwidth}{!}{
\includegraphics{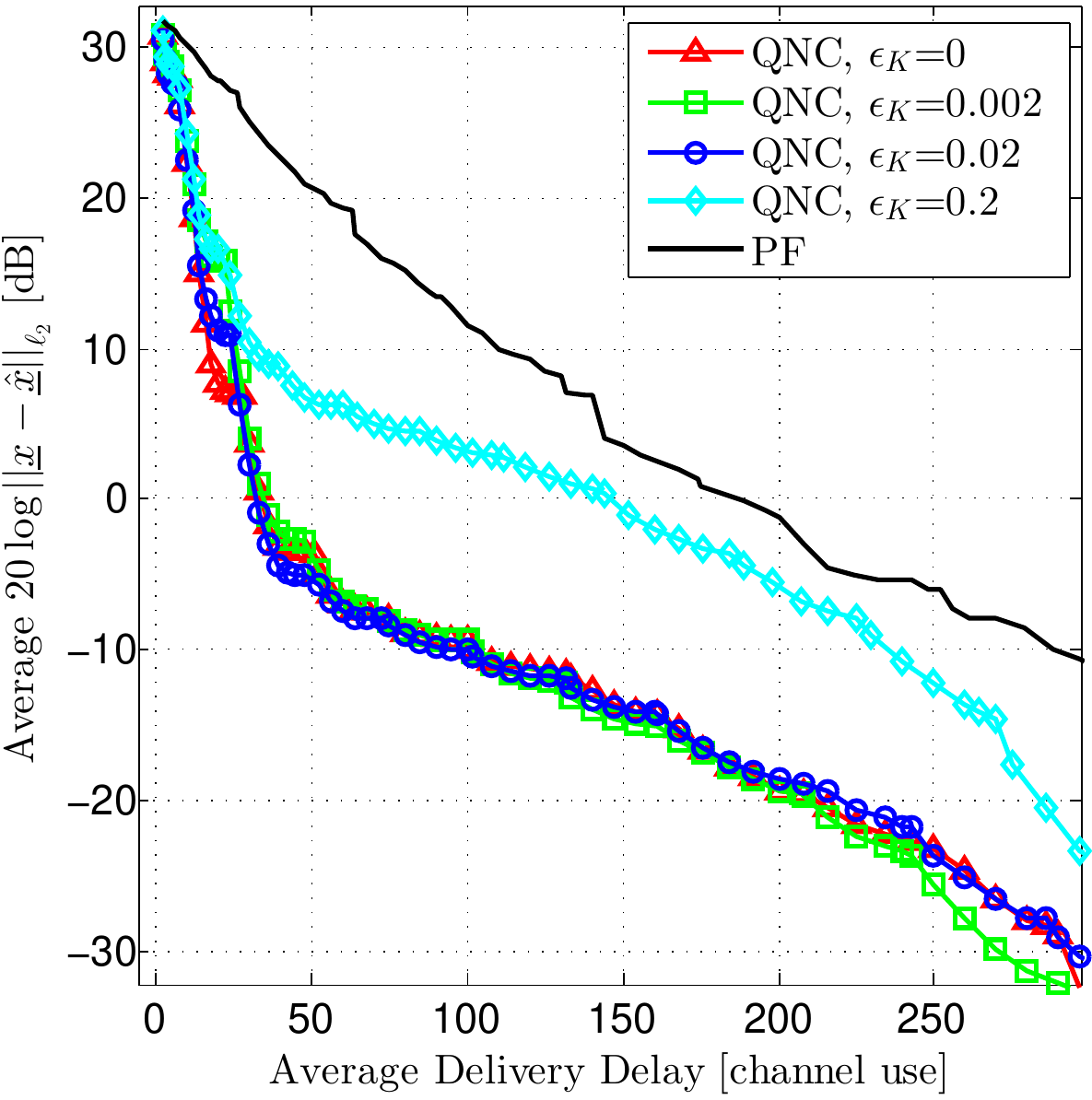}}
\label{fig:1100edgessp1}
} \qquad
\subfigure[$1400$ edges, $\frac{k}{n}=0.15$]{
\resizebox{.41\textwidth}{!}{
\includegraphics{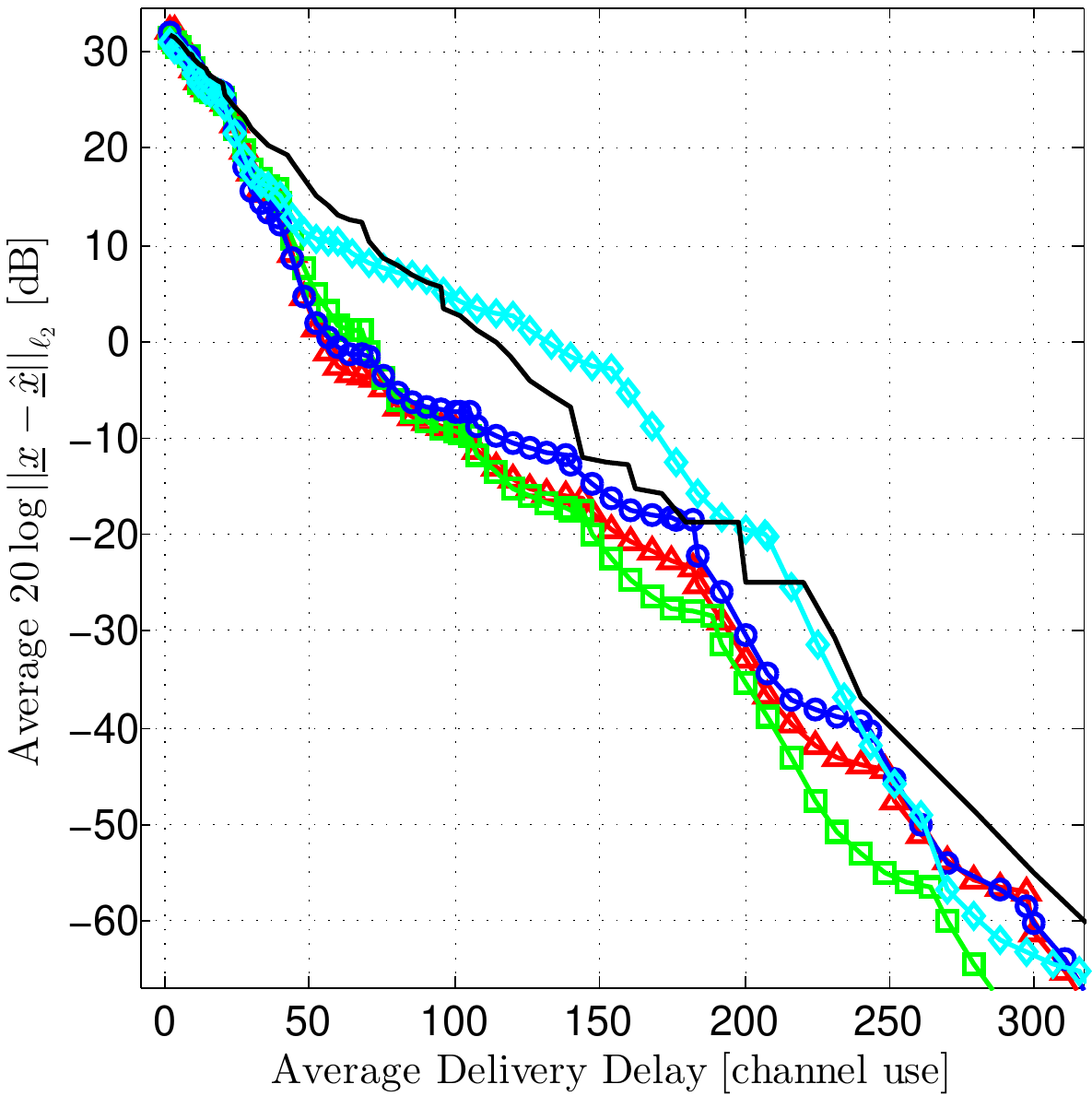}}
\label{fig:1400edgessp2}
} \\
\subfigure[$1800$ edges, $\frac{k}{n}=0.25$]{
\resizebox{.41\textwidth}{!}{
\includegraphics{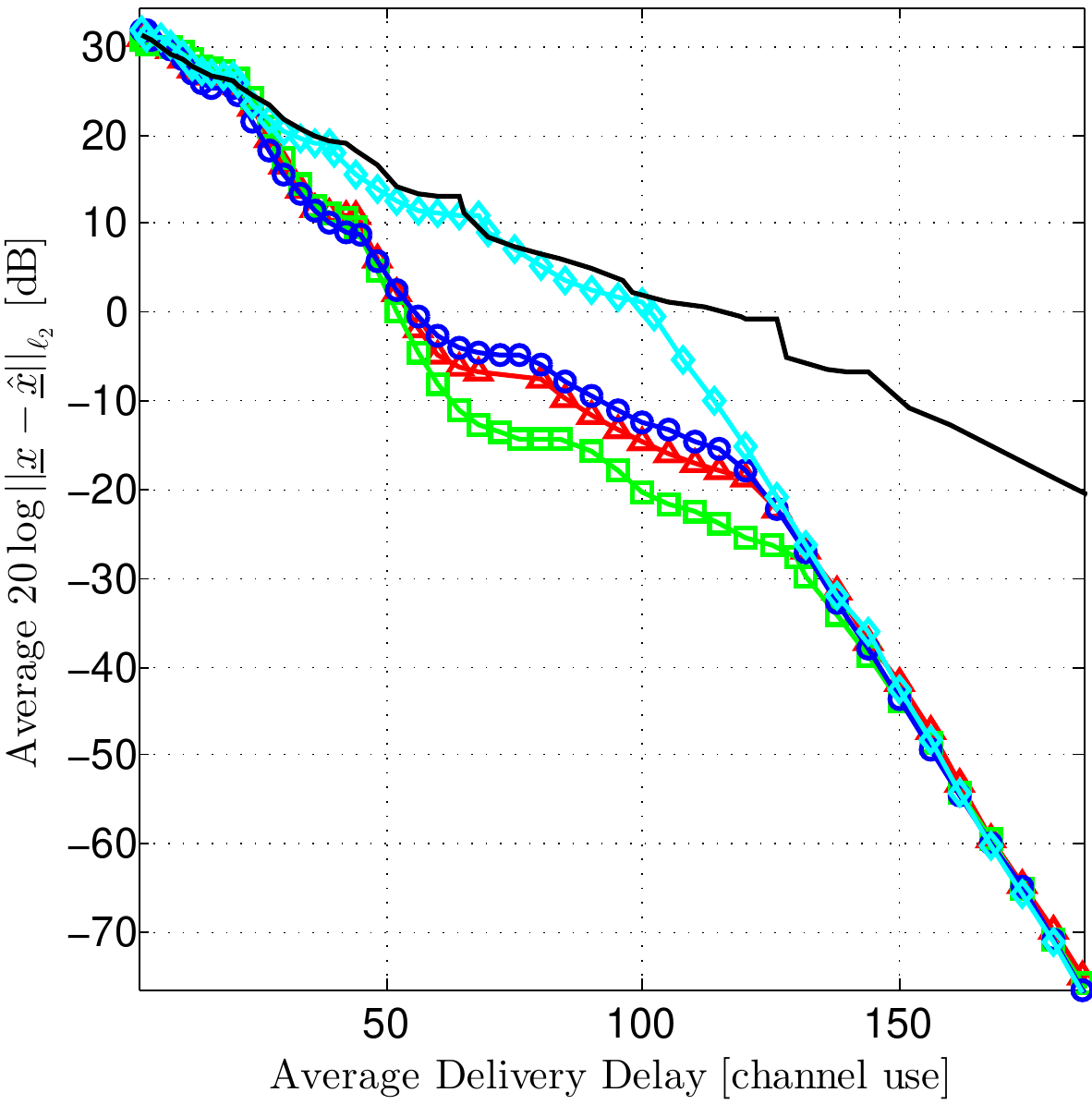}}
\label{fig:1800edgessp3}
} 
\caption{Comparison of QNC versus PF performance for different near-sparsity parameters, $\epsilon_k$ \label{fig:epsKanalysis}.}
\end{figure*}

As a drawback, QNC seems to fail when the sparsity model is not good for describing the correlation model.
Specifically, if the near-sparsity parameter, $\epsilon_k$, is too high, then the resulting performance of QNC can not even achieve that of PF, for a wide range of delivery delays (see Fig.~\ref{fig:1100edgesEpsk4} for instance).
In Figs.~\ref{fig:1100edgessp1},\ref{fig:1400edgessp2},\ref{fig:1800edgessp3}, the effect of $\epsilon_k$ on the resulting QNC performance is illustrated.
As it is shown, as long as $\epsilon_k$ is small (relative to the $\ell_1$ norm of message), there is not any difference in the QNC performance.
But, if it is so high that the sparsity model does not characterize the messages fairly, then QNC fails to work properly (since $\ell_1$ min decoding criteria is not a good cost function anymore).

In the routing based packet forwarding scenarios, the intermediate (sensor) nodes have to go through route training and storage of packets.
As one of the main advantages of network coding, in QNC scenario, the intermediate nodes should only carry simple linear combination and quantization, being liberated in terms of computational complexity.
On the other side, at the decoder sides, QNC requires an $\ell_1$-min decoder which is potentially more complex than the receiver required for packet forwarding.
However, this may not be an issue in practical cases, as the gateway node is usually capable of handling high computational operations.

\section{Conclusions and Future Works}
\label{sec:conclusions}

Joint source network coding of correlated sources was studied with a sparse recovery perspective.
In order to achieve non-adaptive encoding, we proposed quantized network coding, which incorporates real field network coding and quantization to take advantage of decoding using linear programming.
Thanks to the work in the literature of compressed sensing, we discussed theoretical guarantees to ensure efficient encoding and robust decoding of messages.
Moreover, we were able to make conclusive statements about the robust recovery of messages, when fewer number of received packets than the number of sources (messages) were available at the decoder.
Finally, our computer simulations verified the reduction in the average delivery delay, by using quantized network coding.

Although the proposed sparse recovery algorithm is working well for correlated messages with near-sparse characterization, it does not offer optimal recovery for other cases of correlated sources.
Currently, we are studying the feasibility of near-optimal decoding, when other forms of prior information are available about the source.
Specifically, we have suggested the use of belief propagation based decoding \cite{naba3,naba4} in a Bayesian scenario.
However, more theoretical work is needed to derive mathematical guarantees for robust recovery.
Studying the general case of lossy networks with interference between the links is also one of the urging needs for our work.

\bibliographystyle{ieeetr}
\bibliography{ref_arXiv}

\begin{thebibliography}{10}

\bibitem{akyildiz2002survey}
I.~Akyildiz, W.~Su, Y.~Sankarasubramaniam, and E.~Cayirci, ``A survey on sensor
  networks,'' {\em IEEE Communications Magazine}, vol.~40, no.~8, pp.~102--114,
  2002.

\bibitem{chong2003sensor}
C.~Chong and S.~Kumar, ``Sensor networks: evolution, opportunities, and
  challenges,'' {\em Proceedings of the IEEE}, vol.~91, no.~8, pp.~1247--1256,
  2003.

\bibitem{netInfFlow}
R.~Ahlswede, N.~Cai, S.-Y. Li, and R.~Yeung, ``Network information flow,'' {\em
  IEEE Transactions on Information Theory}, vol.~46, pp.~1204 --1216, July
  2000.

\bibitem{al2004routing}
J.~Al-Karaki and A.~Kamal, ``Routing techniques in wireless sensor networks: a
  survey,'' {\em IEEE Wireless Communications}, vol.~11, no.~6, pp.~6--28,
  2004.

\bibitem{slepian1973noiseless}
D.~Slepian and J.~Wolf, ``Noiseless coding of correlated information sources,''
  {\em IEEE Transactions on Information Theory}, vol.~19, no.~4, pp.~471--480,
  1973.

\bibitem{xiong2004distributed}
Z.~Xiong, A.~Liveris, and S.~Cheng, ``Distributed source coding for sensor
  networks,'' {\em IEEE Signal Processing Magazine}, vol.~21, no.~5,
  pp.~80--94, 2004.

\bibitem{SWCtheorem}
T.~S. Han, ``Slepian-wolf-cover theorem for networks of channels,'' {\em
  Information and Control}, vol.~47, no.~1, pp.~67 -- 83, 1980.

\bibitem{1228459}
T.~Ho, R.~Koetter, M.~Medard, D.~Karger, and M.~Effros, ``The benefits of
  coding over routing in a randomized setting,'' in {\em IEEE International
  Symposium on Information Theory}, p.~442, june-4 july 2003.

\bibitem{fragouli2009network}
C.~Fragouli, ``Network coding for sensor networks,'' {\em Handbook on Array
  Processing and Sensor Networks}, pp.~645--667, 2009.

\bibitem{koetter2003algebraic}
R.~Koetter and M.~M{\'e}dard, ``An algebraic approach to network coding,'' {\em
  IEEE Transactions on Networking}, vol.~11, no.~5, pp.~782--795, 2003.

\bibitem{NC_RLNCtoMulticast}
T.~Ho, M.~Medard, R.~Koetter, D.~Karger, M.~Effros, J.~Shi, and B.~Leong, ``A
  random linear network coding approach to multicast,'' {\em IEEE Transactions
  on Information Theory}, vol.~52, no.~10, pp.~4413 --4430, 2006.

\bibitem{lim2011noisy}
S.~Lim, Y.~Kim, A.~El~Gamal, and S.~Chung, ``Noisy network coding,'' {\em IEEE
  Transactions on Information Theory}, vol.~57, no.~5, pp.~3132--3152, 2011.

\bibitem{erasNetCap}
A.~Dana, R.~Gowaikar, R.~Palanki, B.~Hassibi, and M.~Effros, ``Capacity of
  wireless erasure networks,'' {\em IEEE Transactions on Information Theory},
  vol.~52, no.~3, pp.~789 --804, 2006.

\bibitem{ho2004network}
T.~Ho, M.~M{\'e}dard, M.~Effros, R.~Koetter, and D.~Karger, ``Network coding
  for correlated sources,'' in {\em Proceedings of Conference on Information
  Sciences and Systems}, 2004.

\bibitem{NCCorr_SepSCNC}
A.~Ramamoorthy, K.~Jain, P.~A. Chou, and M.~Effros, ``Separating distributed
  source coding from network coding,'' {\em IEEE Transactions on Networking},
  vol.~14, pp.~2785--2795, June 2006.

\bibitem{wu2009practical}
Y.~Wu, V.~Stankovic, Z.~Xiong, and S.~Kung, ``On practical design for joint
  distributed source and network coding,'' {\em IEEE Transactions on
  Information Theory}, vol.~55, no.~4, pp.~1709--1720, 2009.

\bibitem{maierbacher2009practical}
G.~Maierbacher, J.~Barros, and M.~M{\'e}dard, ``Practical source-network
  decoding,'' in {\em 6th International Symposium on Wireless Communication
  Systems}, pp.~283--287, IEEE, 2009.

\bibitem{cruz2011joint}
S.~Cruz, G.~Maierbacher, and J.~Barros, ``Joint source-network coding for
  large-scale sensor networks,'' in {\em IEEE International Symposium on
  Information Theory Proceedings}, pp.~420--424, IEEE, 2011.

\bibitem{kschischang2001factor}
F.~Kschischang, B.~Frey, and H.~Loeliger, ``Factor graphs and the sum-product
  algorithm,'' {\em IEEE Transactions on Information Theory}, vol.~47, no.~2,
  pp.~498--519, 2001.

\bibitem{CS}
D.~Donoho, ``Compressed sensing,'' {\em IEEE Transactions on Information
  Theory}, vol.~52, pp.~1289 --1306, April 2006.

\bibitem{CSbook}
R.~Baraniuk, M.~Davenport, M.~Duarte, and C.~Hegde, {\em An Introduction to
  Compressive Sensing}.
\newblock Addison-Wesley, 2011.

\bibitem{rabbat}
J.~Haupt, W.~Bajwa, M.~Rabbat, and R.~Nowak, ``Compressed sensing for networked
  data,'' {\em IEEE Signal Processing Magazine}, vol.~25, pp.~92 --101, march
  2008.

\bibitem{netcompass}
N.~Nguyen, D.~Jones, and S.~Krishnamurthy, ``Netcompress: Coupling network
  coding and compressed sensing for efficient data communication in wireless
  sensor networks,'' in {\em 2010 IEEE Workshop on Signal Processing Systems},
  pp.~356 --361, oct. 2010.

\bibitem{CdataGathering}
C.~Luo, F.~Wu, J.~Sun, and C.~W. Chen, ``Compressive data gathering for
  large-scale wireless sensor networks,'' in {\em Proceedings of the 15th
  annual international conference on Mobile computing and networking}, MobiCom
  '09, (New York, NY, USA), pp.~145--156, ACM, 2009.

\bibitem{feizi2010compressive}
S.~Feizi, M.~M{\'e}dard, and M.~Effros, ``Compressive sensing over networks,''
  in {\em 48th Annual Allerton Conference on Communication, Control, and
  Computing}, pp.~1129--1136, IEEE, 2010.

\bibitem{xu2011compressive}
W.~Xu, E.~Mallada, and A.~Tang, ``Compressive sensing over graphs,'' in {\em
  IEEE International Conference on Computer Communications (INFOCOM)},
  pp.~2087--2095, IEEE, 2011.

\bibitem{wang2012sparse}
M.~Wang, W.~Xu, E.~Mallada, and A.~Tang, ``Sparse recovery with graph
  constraints: Fundamental limits and measurement construction,'' in {\em IEEE
  International Conference on Computer Communications (INFOCOM)},
  pp.~1871--1879, IEEE, 2012.

\bibitem{feizi2011power}
S.~Feizi and M.~Medard, ``A power efficient sensing/communication scheme: Joint
  source-channel-network coding by using compressive sensing,'' in {\em 49th
  Annual Allerton Conference on Communication, Control, and Computing},
  pp.~1048--1054, IEEE, 2011.

\bibitem{bassi2012compressive}
F.~Bassi, L.~Chao, L.~Iwaza, M.~Kieffer, {\em et~al.}, ``Compressive linear
  network coding for efficient data collection in wireless sensor networks,''
  in {\em Proceedings of the 2012 European Signal Processing Conference},
  pp.~1--5, 2012.

\bibitem{naba1}
M.~Nabaee and F.~Labeau, ``Quantized network coding for sparse messages,'' {\em
  arXiv preprint arXiv:1201.6271}, 2012.

\bibitem{kailath1980linear}
T.~Kailath, {\em Linear systems}, vol.~1.
\newblock Prentice-Hall Englewood Cliffs, NJ, 1980.

\bibitem{candes2007sparsity}
E.~Candes and J.~Romberg, ``Sparsity and incoherence in compressive sampling,''
  {\em Inverse problems}, vol.~23, no.~3, p.~969, 2007.

\bibitem{candes}
E.~J. Candès, ``The restricted isometry property and its implications for
  compressed sensing,'' {\em Comptes Rendus Mathematique}, vol.~346, no.~9-10,
  pp.~589 -- 592, 2008.

\bibitem{baraniuk2007compressive}
R.~Baraniuk, ``Compressive sensing,'' {\em IEEE Signal Processing Magazine},
  vol.~24, no.~4, pp.~118--121, 2007.

\bibitem{simpleProof}
R.~Baraniuk, M.~Davenport, R.~Devore, and M.~Wakin, ``A simple proof of the
  restricted isometry property for random matrices,'' {\em Constr. Approx},
  vol.~2008, 2007.

\bibitem{naba2}
M.~Nabaee and F.~Labeau, ``Restricted isometry property in quantized network
  coding of sparse messages,'' {\em arXiv preprint arXiv:1203.1892}, 2012.

\bibitem{LinProg}
E.~Candes and T.~Tao, ``Decoding by linear programming,'' {\em IEEE
  Transactions on Information Theory}, vol.~51, pp.~4203 -- 4215, December
  2005.

\bibitem{cvx1}
M.~Grant and S.~Boyd, ``{CVX}: Matlab software for disciplined convex
  programming, version 1.21.'' {http://cvxr.com/cvx}, 2011.

\bibitem{cvx2}
M.~Grant and S.~Boyd, ``Graph implementations for nonsmooth convex programs,''
  in {\em Recent Advances in Learning and Control} (V.~Blondel, S.~Boyd, and
  H.~Kimura, eds.), Lecture Notes in Control and Information Sciences,
  pp.~95--110, Springer-Verlag Limited, 2008.

\bibitem{dijkstra1959note}
E.~Dijkstra, ``A note on two problems in connexion with graphs,'' {\em
  Numerische mathematik}, vol.~1, no.~1, pp.~269--271, 1959.

\bibitem{naba3}
M.~Nabaee and F.~Labeau, ``Bayesian quantized network coding via belief
  propagation,'' {\em arXiv preprint arXiv:1209.1679}, 2012.

\bibitem{naba4}
M.~Nabaee and F.~Labeau, ``One-step quantized network coding for near sparse
  messages,'' {\em arXiv preprint arXiv:1210.7399}, 2012.

\end{thebibliography}

\end{document}